\documentclass[a4paper, twoside, fontsize=9pt, twocolumn]{scrartcl}
\pdfoutput=1

\usepackage[a4paper, top=88pt, bottom=88pt, left=50pt, right=50pt, headsep=16pt, footskip=28pt]{geometry}
\usepackage{amsfonts,amssymb,amsmath,amsthm,amstext,amsopn,mathtools,nicefrac,xfrac}
\usepackage[authoryear,sort,round]{natbib} 
\usepackage[utf8]{inputenc}                 
\usepackage{booktabs}                       
\usepackage{nicefrac}                       
\usepackage{microtype}                      
\usepackage{graphics,graphicx}              
\usepackage{subcaption}                     
\usepackage{enumitem}
\usepackage{lipsum}  

\usepackage[boxruled,linesnumbered]{algorithm2e} 
\SetKwComment{Comment}{\%}{}
\SetKwInput{KwInput}{Input}
\SetKwInput{KwOutput}{Output}

\usepackage{verbatim}                       
\usepackage{appendix}                       
\usepackage{bm}                             
\usepackage{array}                          
\newcolumntype{C}[1]{>{\centering\arraybackslash}m{#1}}  
\usepackage[dvipsnames]{xcolor}             
\usepackage{hyperref}                       
\hypersetup{colorlinks=true,linkcolor=Maroon,filecolor=Magenta,urlcolor=Blue,citecolor=RoyalBlue}
\usepackage[noabbrev,capitalise,nosort,nameinlink]{cleveref}  
\usepackage{bbm}                            
\usepackage{mathtools}                      
\usepackage{todonotes}

\newcommand*{\arXiv}[1]{\bgroup\color{blue}\href{https://arxiv.org/abs/#1}{arXiv:#1}\egroup}
\newcommand*{\doi}[1]{\bgroup\color{blue}\href{https://doi.org/#1}{doi:#1}\egroup}
\newcommand*{\email}[1]{\bgroup\color{blue}\href{mailto:#1}{#1}\egroup}
\renewcommand*{\url}[1]{\bgroup\color{blue}\href{#1}{#1}\egroup}
\usepackage{enumitem, moreenum}
\setlist[enumerate]{nosep}
\setlist[itemize]{nosep}
\usepackage{mleftright} \mleftright
\renewcommand{\qedsymbol}{$\blacksquare$}
\renewenvironment{proof}[1][\proofname]{\noindent{\bfseries\sffamily #1.} }{\hfill\qedsymbol\medskip}
\usepackage[textfont={small}, labelfont={sf,bf,small},format=plain,indention=0cm]{caption}
\DeclareCaptionLabelSeparator{figlabelsep}{\,\,\,}
\usepackage{scrlayer-scrpage, xhfill}
\automark[section]{section}
\setkomafont{pageheadfoot}{\normalcolor\sffamily}
\setkomafont{pagenumber}{\normalfont\normalsize\sffamily}
\clearpairofpagestyles
\let\oldtitle\title
\renewcommand{\title}[1]{\oldtitle{#1}\newcommand{\theshorttitle}{#1}}
\newcommand{\shorttitle}[1]{\renewcommand{\theshorttitle}{#1}}
\let\oldauthor\author
\renewcommand{\author}[1]{\oldauthor{#1}\newcommand{\theshortauthor}{#1}}
\newcommand{\shortauthor}[1]{\renewcommand{\theshortauthor}{#1}}
\cohead{\xrfill[0.525ex]{0.6pt}~\theshorttitle~\xrfill[0.525ex]{0.6pt}}
\cehead{\xrfill[0.525ex]{0.6pt}~\theshortauthor~\xrfill[0.525ex]{0.6pt}}
\newcommand{\theabstract}[1]{\par\bgroup\noindent\textbf{\textsf{Abstract}} #1\egroup}
\newcommand{\thekeywords}[1]{\par\smallskip\bgroup\noindent\textbf{\textsf{Keywords.}}\newcommand{\and}{ $\bullet$ } #1\egroup}
\newcommand{\themsc}[1]{\par\smallskip\bgroup\noindent\textbf{\textsf{2020 Mathematics Subject Classification.}}\newcommand{\and}{ $\bullet$ } #1\egroup}
\newcommand*{\affilref}[1]{\ref{affiliation#1}}
\newcommand*{\affiliation}[3]{%
  \begingroup
    \renewcommand{\thefootnote}{\fnsymbol{footnote}}%
    \setcounter{footnote}{0}%
    \footnotetext[#1]{\label{affiliation#2} #3}%
  \endgroup
}
\usepackage{siunitx} 

\numberwithin{equation}{section}
\numberwithin{figure}{section}
\numberwithin{table}{section}

\newcommand*{\defeq}{\coloneqq}

\renewcommand*{\geq}{\geqslant}

\newcommand*{\rd}{\mathrm{d}}

\theoremstyle{definition}

\crefname{assumption}{Assumption}{Assumptions}
\Crefname{assumption}{Assumption}{Assumptions}

\title{The FreeGSNKE Pulse Design Tool (FPDT): a computational framework for evolutive plasma scenario and control design}
\shorttitle{The FreeGSNKE Pulse Design Tool}
\author{
    K. Pentland\textsuperscript{\affilref{UKAEA}} 
    \and
    N. C. Amorisco\textsuperscript{\affilref{UKAEA}} 
    \and
    A. Ross\textsuperscript{\affilref{Hartree}}
    \and
    P. Cavestany\textsuperscript{\affilref{Hartree}}
    \and
    T. Nunn\textsuperscript{\affilref{UKAEA}}
    \and
    A. Agnello\textsuperscript{\affilref{Hartree}}
    \and
    G. K. Holt\textsuperscript{\affilref{UKAEA}}
    \and
    G. McArdle\textsuperscript{\affilref{UKAEA}}
    \and
    C. Vincent\textsuperscript{\affilref{UKAEA}}
    \and
    J. Buchanan\textsuperscript{\affilref{UKAEA}}
    \and
    S. J. P. Pamela\textsuperscript{\affilref{UKAEA}}
}
\shortauthor{K.~Pentland et al.}
\date{\today}

\begin{document}
\maketitle

\affiliation{1}{UKAEA}{United Kingdom Atomic Energy Authority, Culham Campus, Abingdon, Oxfordshire, OX14 3DB, United Kingdom\newline (\email{kamran.pentland@ukaea.uk})}
\affiliation{2}{Hartree}{STFC Hartree Centre, Sci-Tech Daresbury, Keckwick Lane, Daresbury, Warrington, WA4 4AD, United Kingdom}


    
\begin{abstract}\small
    \theabstract{%
    We present the FreeGSNKE Pulse Design Tool (FPDT), an open-source, Python-based computational framework that enables in silico testing and predictive design of tokamak plasma scenarios and control strategies.
    The FPDT couples the FreeGSNKE evolutive equilibrium solver with a virtual Plasma Control System (PCS) containing modular and customisable controllers. Given a set of user-defined waveforms and control parameters, the virtual PCS uses feedback and feedforward control to modulate plasma current, position, and shape, while adhering to machine safety limits on poloidal field coil currents and voltages. The resulting framework allows simulation of the controlled dynamic evolution of plasma equilibria, along with the currents in both active poloidal field coils and passive conducting structures, under the assumption of axisymmetry.
    The FPDT can be used to develop plasma scenarios, test control schemes, calibrate control parameters, and perform uncertainty quantification studies, thereby reducing iterative and expensive experimental testing on a physical tokamak. The FPDT is machine-agnostic and can be customised to implement different control algorithms tailored to the specific tokamak of interest.
    Here, we outline the overall framework and validate its performance on plasma discharges on the MAST Upgrade tokamak in the `flat-top' phase. 
    We demonstrate excellent quantitative agreement between the FPDT simulations, the desired control waveforms, and the experimental shot data.
    With this extension to the FreeGSNKE open-source suite of codes we aim to encourage more reproducible and collaborative research in plasma modelling and control.
    }

    \thekeywords{%
        {Pulse design tool}%
        \and%
        {Plasma control system}%
        \and%
        {Grad--Shafranov}%
        \and%
        {MHD equilibria}%
        \and%
        {FreeGSNKE}%
        \and%
        {MAST-U}%
    }
\end{abstract}
 

\section{Introduction} \label{sec:intro}


In magnetic confinement fusion, plant operators must be able to rely on robust control algorithms and machine protection systems in order to safely operate, sustain, and reproduce high-performance tokamak plasma scenarios \citep[e.g.][]{ariola2016}.
To do this, all tokamak devices are equipped with a Plasma Control System (PCS), a combination of software and hardware that receive measurements from diagnostics around the machine in order to actively control both the plant and the plasma using a number of different actuators \citep[see, e.g.,][]{penaflor2008}.
At a minimum, the role of the PCS is to maintain the desired plasma position, shape, and current throughout all phases of a plasma ``pulse''\footnote{Sometimes referred to as a ``discharge'' or ``shot''.}---from coil magnetisation and current ramp-up, through flat-top operation, to ramp-down and termination. 
Beyond magnetic control, the PCS can also be responsible for coordinating systems that regulate plasma heating (e.g. via neutral beams), manage heat loads in the divertor, maintain fuel levels through gas puffing, and monitoring impurities and radiation.
Some example PCSs include those at DIII-D \citep{penaflor2013}, TCV \citep{galperti2024}, and MAST-U \citep{mcardle2020,mcardle2023}.

To gain deeper insight into the interaction between control actuators and plasma dynamics without physically running experiments, plasma experimentalists are increasingly turning to Pulse Design Tools (PDTs).
A PDT is a computational framework for simulating the time-dependent evolution of a tokamak plasma subject to the actions of a (virtual) PCS. 
As such, PDTs enable in silico control design and validation. 
Access to a PDT can support researchers in planning tokamak experiments by enabling simulations of challenging new scenarios and control algorithms, dramatically reducing the need for direct and potentially error-prone testing on the tokamak. 
PDTs can be used to test if a given discharge might violate machine safety limits (e.g. coil currents and voltages), as well as calibrate and optimise any parameters defining the control algorithms (e.g. gains, virtual circuits, damping, etc.) across different operating scenarios and before deployment on the real PCS.
They can support uncertainty quantification and its impact on PCS performance, for example, by simulating how noise and disturbances affect measurement quantities from various diagnostics (whether synthetic or not).
PDTs can also be used for control room training and, if directly integrated with a PCS, can enable automated pulse testing and execution---reducing the chance of human error in the experimental pipeline.

In this paper, we focus on the magnetic control of the plasma position, shape, and current, via their actuators---the poloidal field (PF) coils.
The control actions consist of active voltages applied by the power supplies to the PF coils.
These voltages generate currents, which in turn produce magnetic fields that confine, shape, and stabilise both the core plasma (keeping it away from the vessel walls) and the divertor legs (managing exhaust heat loads on the divertor tiles).
Using diagnostic measurements, the PCS continually adjusts these voltages in real-time (on sub-\si{\milli\second} timescales) such that user-prescribed waveforms for plasma current, shape, and position are followed, thus maintaining a stable operating scenario.
In the context of magnetic control, PDTs typically include two main components: (i) a fast\footnote{Relative to the time between plasma discharges.}, accurate evolutive simulator of the coupled plasma, PF coil, and vessel currents, and (ii) a virtual PCS composed of several interlinked control algorithms.
A given PCS can employ a wide range of different control algorithms, including proportional–integral–derivative (PID) controllers, model predictive control \citep{gerksic2016,mele2025}, and reinforcement-learning agents \citep{degrave2022,kerboua2024,tracey2024}.

A vast array of evolutive free-boundary equilibrium simulators have been presented in the literature, varying in programming language, modelling assumptions, implementation choices, computational cost, open accessibility, and ease of use.
An incomplete list includes: FreeGSNKE \citep{amorisco2024}, NICE \citep{gros2025}, CREATE-NL \citep{albanese2015}, GSevolve \citep{welander2019}, DINA \citep{khayrutdinov1993}, SPIDER \citep{ivanov2009}, CEDRES\texttt{++} \citep{heumann2015}, FGE \citep{carpanese2021, heis2025}, and NSFsim \citep{clark2025}.
Similarly, a number of PDTs have been developed and successfully used to support experiments on several tokamaks. 
A few notable examples are: XSC Tools, used routinely on JET \citep{albanese2005,tommasi2007}; TokSys, used on DIII-D \citep{humphreys2008}, MAST-U \citep{anand2024,lvovskiy2025}, and NSTX-U \citep{pajares2025}; MEQ, used on TCV \citep{carpanese2021}; Fenix, deployed on ASDEX-U \citep{fable2022,janky2021}; the Plasma Control System Simulation Platform to be deployed on ITER \citep{walker2015}; SOPHIA, used on ST40 \citep{janky2026}; and NICE, applied to WEST \citep{gros2025,nouailletas2025}.
To the best of our knowledge, NICE\footnote{\url{https://gitlab.inria.fr/blfauger/nice}} and MEQ\footnote{\url{https://gitlab.epfl.ch/spc/public/meq}} are the only open-source PDTs currently available to the community (with the caveat that the MATLAB code requires a licence).

These tools vary widely in their capabilities, from utilising physics models of varying complexity to offering advanced graphical or web-based user interfaces. 
Underlying physics models range from equilibrium-only descriptions (e.g. XCS Tools, TokSys, MEQ) to those that include current diffusion and transport models (e.g. Fenix, SOPHIA, NICE).
Some tools are designed to reflect the physical setup, control, and diagnostic capabilities of a specific tokamak, while others can be suitably modified for use on different machines.
In addition, certain tools employ linear or piecewise-linear model equations to reduce computational runtime and can interface directly with a physical PCS on particular tokamaks (e.g. XCS Tools, TokSys).
Most PDTs focus on predictive simulations, however, inverse evolutive equilibrium tools for scenario design, such as CEDRES\texttt{++} \citep{heumann2015}, FEEQS \citep{blum2019}, and GSPulse \citep{wai2025}, have also been proposed.
We also note that integrated modelling tools---which couple physics, engineering, and control models to design and optimise entire fusion devices rather than individual plasma pulses---do exist, but they are not designed for ``virtual PCS'' shot design and validation simulations.
Examples include PROCESS \citep{muldrew2020}, Bluemira \citep{bluemira}, METIS \citep{artaud2018}, JINTRAC \citep{romanelli2014}, and FUSE \citep{meneghini2024}.

For a PDT to be useful, fundamental properties must include numerical fidelity, speed, and ease of customisation---all of which can make or break its more widespread adoption by the modelling community.
For instance, ensuring that the PDT returns accurate re-simulations of past discharges (assuming it is provided with the corresponding PCS settings) is fundamental when looking to use the PDT to explore new scenarios or modifications on the default control algorithms. 
Computational speed is clearly advantageous, but this will vary depending on the level of physics captured by the simulator and the intended use cases, with control room use for rapid testing between plasma discharges likely being the most demanding. 
In order to simplify design and optimisation, control parameters and waveforms should be easily accessible and editable by users, while control algorithms should also be interchangeable and easily modifiable to accommodate different machines. 

With this in mind, we present here the FreeGSNKE Pulse Design Tool (FPDT), a Python-based PDT built on the evolutive equilibrium code FreeGSNKE \citep{amorisco2024}, and released as part of the FreeGSNKE open-source suite\footnote{\url{https://github.com/FusionComputingLab/freegsnke}}.
The FPDT aims to enable in silico predictive design of experimental plasma scenarios and control strategies.
To do this, we couple FreeGSNKE’s existing evolutive equilibrium solver with a new virtual PCS class that contains modular, documented, and customisable algorithms, each responsible for enforcing different aspects of magnetic plasma control.
This extends FreeGSNKE's previous capability of purely feedforward (FF) equilibrium simulations of plasma pulses that use time-dependent PF coil voltages recorded from, for example, past discharges or other sources. 
The FPDT now enables fully flexible control of plasma current, position, and shape, in which voltage actuators are driven by a combination of both FF and feedback (FB) controllers.
It has a modular and transparent design, which supports implementation of different internal control algorithms tailored to specific tokamak configurations, and could potentially serve as the foundation of a more comprehensive ``flight simulator'' tool for plasma modelling.

One of the motivations for developing this tool is to provide the community with an open-source PDT built on a contemporary software stack.
The FPDT is free and open-source, with comprehensive documentation and fully accessible source code that enables scrutiny and community contributions.
We hope the FPDT will enable more reproducible and collaborative research in plasma modelling and control studies \citep{gsfit2025,farmakalides2025, sarkar2026, kuang2026}, particularly when developing modern control approaches based on machine learning, with which FreeGSNKE integrates seamlessly \citep{pentland2026}.

While the FPDT is inherently machine agnostic, the controllers we implement in the virtual PCS shown here take inspiration from those in the MAST-U PCS \citep{mcardle2020,mcardle2023}, which have been successfully used to demonstrate high-performance scenarios with highly shaped plasmas, advanced divertor configurations, and heat flux mitigation \citep{morris2018, kool2025}.
These include PID-based algorithms to control the plasma current, vertical plasma position, and both local (e.g. midplane radii, gaps, X-point positions) and global (e.g. elongation, triangularity, squareness) shape parameters and 
are accompanied by algorithms to handle PF coil activation times and safety constraints (in the form of current and voltage limits). 
We note that we do not include actuators and subsystems required to regulate plasma heating, fuelling, impurities, radiation, and detachment. 

Throughout the rest of the paper, we demonstrate how the FPDT satisfies the aforementioned criteria for an effective PDT and more specifically we
\begin{enumerate}[label=(\roman*)]
    \item describe the FPDT and the coupling between FreeGSNKE’s evolutive equilibrium solver and the new PCS class (\cref{sec:overview}).
    \item detail the governing equations of the simulator (\cref{sec:simulator}) and the implementation of the control algorithms within the PCS class, linked to MAST-U (\cref{sec:PCS}).
    \item validate the FPDT simulations against challenging plasma scenarios from previously executed MAST-U pulses (\cref{sec:results}). 
\end{enumerate}
We discuss the implications of our work in \cref{sec:discussion} and conclude by outlining the future of the FPDT.

\section{The FreeGSNKE Pulse Design Tool} \label{sec:PDT}

\subsection{Overview} \label{sec:overview}
\begin{figure}[t!]
    \begin{subfigure}{0.99\linewidth}
        \centering
        \includegraphics[width=0.99\textwidth]{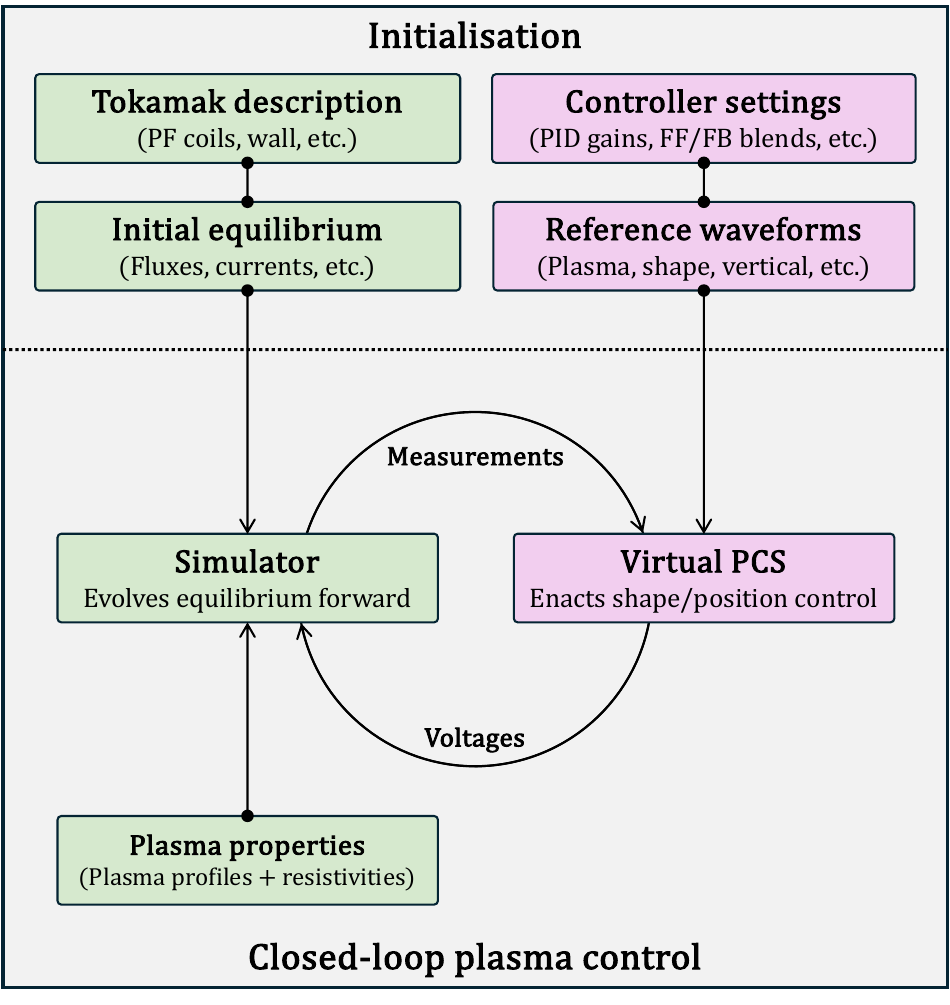}
    \end{subfigure}
    \caption{High-level schematic of how an FPDT simulation operates.
    }
    \label{fig:PDT_schematic_high_level}
\end{figure}

At a high level (see the schematic in \cref{fig:PDT_schematic_high_level}), the FPDT simulates the dynamic evolution of plasma equilibria and the associated currents in both the active PF coils and passive conducting structures using voltages generated by a virtual PCS class.
In addition to the tokamak description and a set of initial conditions (i.e. a starting equilibrium), the FPDT requires user-defined reference \emph{waveforms} for key plasma parameters (e.g. plasma current, vertical position) and controller settings that will govern the desired scenario.

During an FPDT simulation, FreeGSNKE's evolutive solver and the virtual PCS class interact during each time stepping cycle.
At time $t$, the FreeGSNKE plasma state (see \cref{sec:plasma_state}) is queried and provides the virtual PCS with simulated measurements of the total plasma current, PF coil currents, and a number of plasma shape parameters.
Based on the user-defined control settings and reference waveforms, these measurements are used to calculate the voltages that should be applied to the PF coils.
Using these voltages and some prescribed internal plasma properties, FreeGSNKE evolves the plasma state from time $t$ to $t+dt$, at which point the PCS is once again provided updated measurements, starting a new cycle. 
This closed loop is repeated until the desired end of the simulation, with users being able to extract any equilibrium-related quantities of interest at each time step.


\subsection{The FreeGSNKE simulator} \label{sec:simulator}
Here, we provide a brief overview of FreeGSNKE, its governing equations, and how it works, focusing on aspects directly relevant to the FPDT---for a more complete description, we refer the reader to \citep{amorisco2024}.

\subsubsection{Governing equations} \label{sec:evolution}
The governing equations in FreeGSNKE describe the evolution of the plasma state together with the currents flowing in any conducting metal structures surrounding the tokamak. 
This is achieved by coupling the nonlinear Grad–Shafranov (GS) equation \citep{grad1958,shafranov1958} with a set of time-dependent circuit equations that relate the currents in the conducting metals to the currents in the plasma itself.

Firstly, the circuit equation governing the flow of current in the PF coils and any conducting passive structures at time $t$ is given by
\begin{align} \label{eq:circuit_eqs}
    \mathcal{M}_{m,m} \frac{\rd}{\rd t} \bm{I}_m + \mathcal{M}_{m,y} \frac{\rd}{\rd t} \bm{I}_y + \mathcal{R}_{m,m} \bm{I}_m = \bm{V}_m \ ,
\end{align}
where
\begin{itemize}
    \item $\bm{I}_{m}(t) = (\bm{I}_\text{act}(t), \bm{I}_\text{pas}(t))$ is the vector of metal currents, including both active PF coils\footnote{In this work, active PF coils refers to all actively controlled coils, including Ohmic coils, shape control coils, and vertical control coils.} and passive conducting structures.
    \item $\bm{I}_{y}(t)$ is the vector describing the discretised current distribution of the plasma on the solution domain, which (in FreeGSNKE) is restricted to the region inside the plasma limiter (i.e. the first wall).
    \item $\bm{V}_{m}(t) = (\bm{V}_\text{act}(t), \bm{0})$ is the vector of voltages applied to the metal elements, comprising the active PF coils (whose voltages serve as the control actions supplied by the virtual PCS) and the passive structures (which receive no active voltage).
    \item $\mathcal{M}_{m,m}$ and $\mathcal{R}_{m,m}$ denote the mutual inductance and (diagonal) resistance matrix relevant to the metals, respectively. 
    $\mathcal{M}_{m,y}$ denotes the inductance matrix of all metals on the discretised plasma  domain\footnote{We use bold symbols to represent vectors and calligraphic uppercase letters to denote matrices.}.
\end{itemize}

Secondly, to describe the plasma, FreeGSNKE uses a ``lumped'' one-dimensional equation given by
\begin{align} \label{eq:plasma_eq}
   \frac{\bm{I}_y^{\intercal}}{I_p} \cdot \left[ \mathcal{M}_{y,y} \frac{\rd}{\rd t} \bm{I}_y + \mathcal{M}_{m,y}^{\intercal} \frac{\rd}{\rd t} \bm{I}_m + \mathcal{R}_{y,y} \bm{I}_y \right] = 0 \ ,
\end{align}
where
\begin{itemize}
    \item $I_p(t)$ is the total plasma current.
    \item $\mathcal{R}_{y,y}(t)$ denotes the (diagonal) resistance matrix of the plasma.
    We define
    \begin{align*}
    \text{diag} \ \mathcal{R}_{y,y}(t) = 2 \pi (\bm{R}_{y} \oslash \bm{A}_{y}) \rho_p(t),
    \end{align*}
    with $\bm{R}_y$ being the cylindrical radii of each domain element, $\bm{A}_y$ their areas, and $\rho_p(t)$ the (effective) user-defined plasma resistivity---the $\oslash$ denotes element-wise division. 
    \item $\mathcal{M}_{y,y}$ is the inductance matrix of plasma-to-plasma current elements.
\end{itemize}

Finally, the GS equation, which describes the magnetohydrodynamic equilibrium (force balance) of an axisymmetric plasma, is given by 
\begin{align}
    \Delta^* \psi(R,Z) &= -\mu_0 R J_{\phi}(\psi, R, Z; \bm{\theta}), \quad &&(R,Z) \in \Omega, \label{eq:GS_PDE} 
\end{align}
expressed in cylindrical coordinates $(R,\phi,Z)$ over the rectangular domain $\Omega$.
Here, we have the linear elliptic operator $\Delta^* = \defeq R \partial_R R^{-1} \partial_R + \partial_{ZZ}$, the poloidal magnetic flux $\psi$, the magnetic permeability of free space $\mu_0$, and the (toroidal) plasma current density distribution $J_\phi$.
This equation is coupled with a Dirichlet boundary condition on the boundary ($\partial \Omega$) of the computational domain---see \citet[][eq.~3.4]{pentland2024}.

The vector $\bm{\theta}$ contains the ``profile parameters'' that define the form of the  plasma current density distribution $J_{\phi}$ inside the last closed flux surface (LCFS) of the plasma, in addition to the safety factor $q$.
In particular, they define the shapes of $\rd p / \rd \psi$ and $F \rd F / \rd \psi$, where $p(\psi)$ denotes the pressure and $F(\psi)$ denotes the toroidal magnetic field profile.
While many different parametrisations exist, we adopt the \cite{lao1985} polynomial descriptions---more details on this implementation can be found in \cite{pentland2024}.

\subsubsection{Evolved tokamak plasma state and user-defined quantities} \label{sec:plasma_state}
By coupling and solving equations \eqref{eq:circuit_eqs}-\eqref{eq:GS_PDE}, FreeGSNKE determines the evolution of the tokamak plasma state
\begin{align*}
    \bm{S}(t) = \Big( \underbrace{\bm{I}_m(t), {I}_p(t)}_{\text{evolved}} \ ; \ \underbrace{\bm{\theta}(t), \rho_p(t)}_{\text{prescribed}} \Big),
\end{align*}
ensuring the plasma proceeds through configurations that remain in equilibrium on the Alfv\'{e}n timescale \citep{jardin2010}.
The plasma equilibrium at time $t$, given through the GS equation, is fully described by the set of metal currents $\bm{I}_m(t)$, the total plasma current ${I}_p(t)$, and the profile parameters $\bm{\theta}(t)$.
While these quantities are sufficient for describing the instantaneous plasma state, the resistivity $\rho_p(t)$ is also required to define evolution.

The evolution of both $\bm{I}_m(t)$ and $I_p(t)$ are fully determined in FreeGSNKE by \eqref{eq:circuit_eqs}-\eqref{eq:GS_PDE}. 
However, the user needs to fully define the time evolution of both $\bm{\theta}(t)$ and $\rho_p(t)$ for the duration of the simulated time interval. 
This is because, at present, FreeGSNKE does not directly model current diffusion or self-consistently evolve the resistivity within the plasma---these features will be introduced in the future.
From the tokamak plasma state $\bm{S}(t)$, we can derive required quantities of interest, including any simulated measurements needed in the virtual PCS at each time step.
These measurements enable the virtual PCS to provide the PF coil voltages $\bm{V}_\text{act}(t)$ back to the simulator, which are then used in the circuit equations \eqref{eq:circuit_eqs} for evolution to the next time step---recall \cref{fig:PDT_schematic_high_level}.

Before moving forward, we reiterate that $I_p(t)$ is evolved self-consistently by FreeGSNKE rather than prescribed through the assigned plasma profile parameters $\bm{\theta}(t)$.
These parameters determine the \emph{shape} of the $\rd p / \rd \psi$ and $F \rd F / \rd \psi$ profiles, as well as their \textit{relative} size, whereas their \emph{absolute} size is set by $I_p(t)$ itself.

\subsubsection{Solution modes}
The FPDT offers three different solution modes for the system \eqref{eq:circuit_eqs}-\eqref{eq:GS_PDE}, namely:
\begin{itemize}
    \item Fully nonlinear evolution; which we refer to with NL.
    \item Piecewise linear evolution of the dynamics \eqref{eq:circuit_eqs}-\eqref{eq:plasma_eq} coupled with fully nonlinear solution of GS \eqref{eq:GS_PDE}; which we refer to with PwLD.
    \item Piecewise linear evolution for the dynamics \eqref{eq:circuit_eqs}-\eqref{eq:plasma_eq} and GS \eqref{eq:GS_PDE}; which we refer to with PwL.
\end{itemize}
A full description of the NL mode is provided in \citet[Sec. IV B]{amorisco2024}.
It uses an implicit Euler scheme to discretise \eqref{eq:circuit_eqs}-\eqref{eq:plasma_eq} in time and a staggered approach, based on the Newton–Krylov method, to solve the resulting nonlinear system with \eqref{eq:GS_PDE}. 
Similarly, the linearisation of \eqref{eq:circuit_eqs}-\eqref{eq:plasma_eq} used by both the PwLD and PwL modes is presented in \citet[eqs.~28 and~29]{amorisco2024}.
It relies on (finite difference) Jacobians of the plasma current distribution $\bm{I}_{y}$ with respect to all of the independent variables at some linearisation time $t_0$: the metal currents $\bm{I}_m$, plasma current $I_p$, and profile parameters $\bm{\theta}$.
Here, we introduce the ability to automatically update these Jacobians as the simulation progresses, which gives rise to the ``piecewise'' aspect. 

Before discussing the criteria used to trigger a re-calculation of the Jacobians, it is useful to clarify the difference between PwLD and PwL modes. 
In the PwLD mode, only the dynamics \eqref{eq:circuit_eqs}-\eqref{eq:plasma_eq} are linearised, meaning that the nonlinear GS equation \eqref{eq:GS_PDE} is solved at every time step (based on metal currents and plasma currents obtained from the evolution of the linearised dynamics equations).
Solution of the GS equation provides the virtual PCS with exact plasma descriptors for each equilibrium, for instance, vertical position and plasma shape parameters. 
In the PwL mode, the GS problem is also linearised, meaning that the observable plasma descriptors necessary for control are not exact but are linearised with respect to the metal currents $\bm{I}_m$, plasma current $I_p$, and profile parameters $\bm{\theta}$ (see \cref{app:PwLD} for further details). 
Both modes are significantly less computationally demanding than the NL mode, however, their accuracy depends on the criteria used to trigger an update of the linearisations. 
In the PwLD mode, we use a criterion based on the evolution of the plasma current distribution: the linearisation is updated when the relative difference (in norm) between the evolving plasma current $\bm{I}_{y}(t)$ and that at the latest linearisation $\bm{I}_{y}(t_0)$ exceeds some threshold. 
In PwL mode, the full plasma current distribution $\bm{I}_{y}(t)$ is not available, as the GS equation is not solved at all time steps. 
Therefore, it requires a criterion based on the relative change in the user-defined plasma descriptors. 
The performance of these criteria is described in \cref{sec:results}.

\subsubsection{Tokamak description, vessel mode reduction, and time step selection}

To initialise the FPDT, a tokamak geometry must be specified. 
At minimum this must include a set of PF coils and a limiter geometry that defines the region accessible to the plasma, and may also include passive conducting structures and diagnostics---see \cite{pentland2024}, \cite{lvovskiy2025}, and the FreeGSNKE documentation for more details. 
When building the tokamak object, FreeGSNKE calculates $\mathcal{M}_{m,m}$ and $\mathcal{R}_{m,m}$ based on the input geometry data and resistivity values specified for the PF coils and passive structures.
The user can override these calculations and provide custom resistance and mutual inductance matrices, which is often necessary when attempting a quantitative comparison with experiment\footnote{In practice, purely geometric resistance and inductance values are often insufficient because, for example, they do not account for wiring between PF coils and the power supplies.}.
Inductances between plasma elements and the PF coils, i.e. $\mathcal{M}_{y,y}$ or $\mathcal{M}_{y,m}$, are calculated based on geometry.

As detailed by \citet[Sec. III A]{amorisco2024}, passive structure eigenmodes that evolve on exceedingly fast timescales can be omitted from the simulation to reduce computational runtime.
It should be noted, however, that retaining fewer modes systematically weakens the passive vertical stabilisation provided by eddy currents and therefore reduces the vertical instability timescale associated with the equilibrium.
As a consequence, the minimum viable number of modes depends on the regime being simulated; therefore, highly elongated plasmas, for instance, may require a higher number of eigenmodes to be retained.\footnote{We refer the reader to the FreeGSNKE documentation for further details: \url{https://docs.freegsnke.com/}.}.
The characteristic timescale of the vertical instability is also crucial to the selection of the simulation time step $dt$. 
In order to resolve the vertical dynamics, the time step should be at least a factor $5\text{–}10$ smaller than the vertical instability timescale, which is calculated automatically by FreeGSNKE's evolutive solver upon initialisation. 

Finally, in order to run the FPDT, the evolutive solver should be provided with an initial condition.
In practice, one typically needs to define a full tokamak-plasma state $\bm{S}(t)$ at $t=t_{\text{init}}$, to allow FreeGSNKE to solve the associated static forward GS problem.
When using the FPDT to simulate a variation on a previous discharge, this initial state can, for example, be built using equilibrium reconstruction data.
\begin{figure*}[t!]
    \begin{subfigure}{0.99\linewidth}
        \centering
        \includegraphics[width=0.9\textwidth]{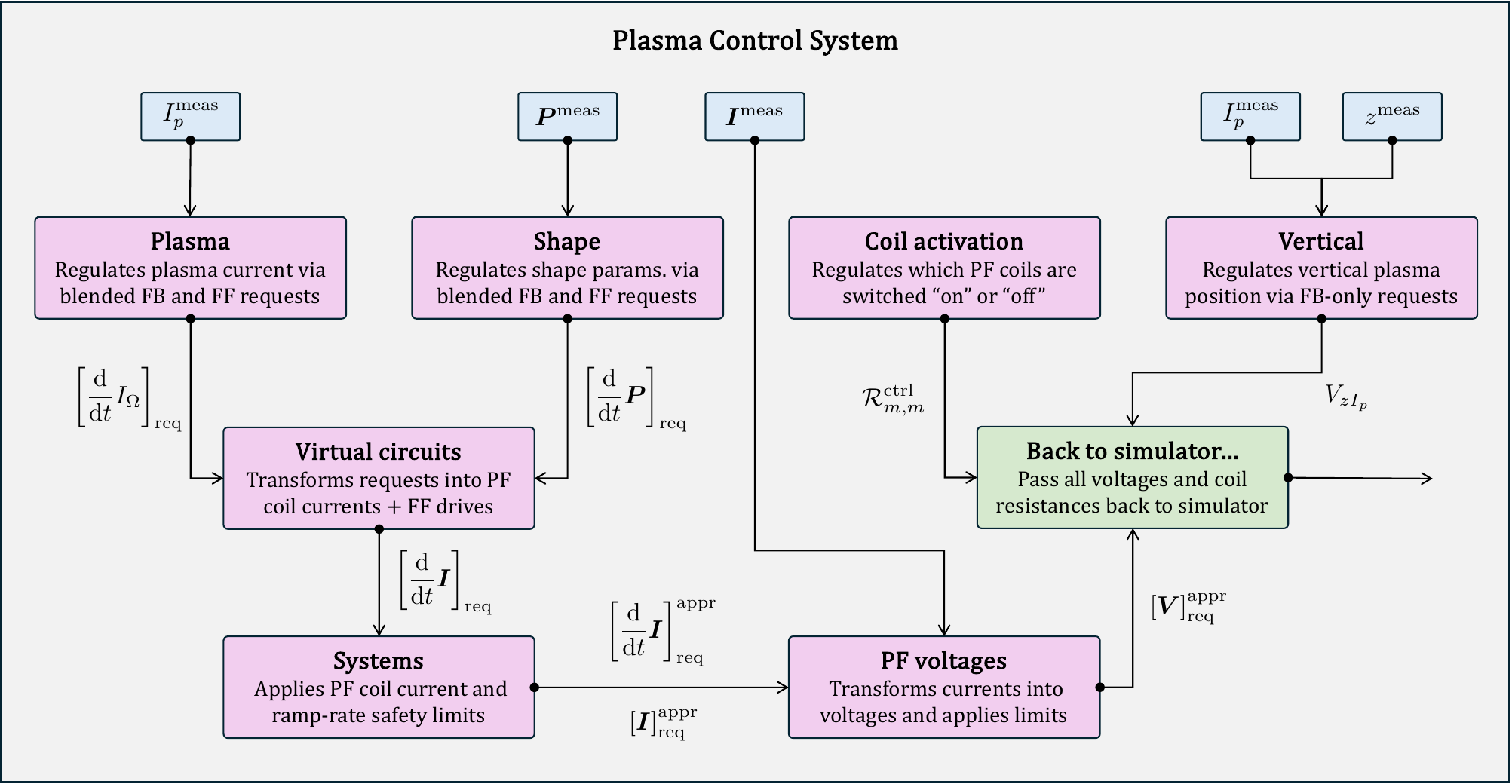}
    \end{subfigure}
    \caption{
    Overview of the virtual PCS in the FPDT.
    Shown are the measurements provided by the simulator (blue), the individual controllers (pink), and the main quantities that pass between each of them---full details of which are given in \cref{sec:simulator,sec:PCS}, respectively.
    }
    \label{fig:PDT_schematic}
\end{figure*}

\subsection{The plasma control system} \label{sec:PCS}

The PCS class in FreeGSNKE is a collection of interchangeable and adaptable algorithms for controlling key plasma quantities via a combination of FF and FB control---see \cref{fig:PDT_schematic} for an illustration of the workflow. 
This modular structure and the customisability of each of the individual controllers makes it straightforward to adapt the system to different control requirements and different machine configurations. 

\subsubsection{FB and FF control}

Before describing each controller separately, we briefly clarify our nomenclature regarding FB and FF control.
In the FPDT, controllers use a combination of both FB and FF requests. 
For a generic quantity $X = X(t)$, we define the total request as
\begin{align} \label{eq:fb_ff}
    \left[\frac{\rd}{\rd t} X\right]_{\text{req}} = b_X \left[\frac{\rd}{\rd t} X\right]_{\text{FB}}  + (1 - b_X) \left[\frac{\rd}{\rd t} X\right]_{\text{FF}} ,
\end{align}
where $b_X = b_X(t) \in [0,1]$ is a ``blend'' coefficient defining the convex combination between the FB and FF terms. 

Throughout the FPDT, the FB terms (as in \eqref{eq:fb_ff}) are provided by PID controllers:
\begin{align}\label{eq:PID}
    \left[\frac{\rd}{\rd t} X\right]_{\text{FB}} = K^{\text{prop}}_X \delta_X + K^{\text{int}}_X \iota_X + K^{\text{der}}_X\frac{\rd}{\rd t}\delta_X,
\end{align}
which correct departures from a desired reference value. 
The proportional term is based on the error 
\begin{align}\label{eq:fbref}
    \delta_X (t) =  X^{\text{ref,FB}} (t) - X^{\text{meas}} (t),
\end{align}
and has gain $K^{\text{prop}}_X = K^{\text{prop}}_X(t)$.
The user-defined reference waveform for $X$ is denoted $X^{\text{ref,FB}} (t)$ and its real-time measurement is $X^{\text{meas}} (t)$.
The integral term, assuming the simulation and PCS time step $\rd t$ are the same here, uses $\iota_X$ as an approximation of the integral of the error:
\begin{align}
     \iota_X (t) =  \iota_X (t-\rd t) + \delta_X(t) \ \rd t,
\end{align}
and has gain $K^{\text{int}}_X =K^{\text{int}}_X (t)$.
\begin{align}
     \frac{\rd}{\rd t}\delta_X (t) =  \big( \delta_X (t) - \delta_X(t-\rd t)\big) / \rd t,
\end{align}
and has gain is $K^{\text{der}}_X =K^{\text{der}}_X (t)$. 

The FF term in \eqref{eq:fb_ff} makes no reference to real-time measurements and enacts prescribed requests which are entirely user-defined. 
\begin{align}\label{eq:FFref}
    \left[\frac{\rd}{\rd t} X\right]_{\text{FF}} = \frac{\rd}{\rd t} X^{\text{ref,FF}} ,
\end{align}
or by setting the FF derivative on the right hand side directly with a waveform.
Note that the FB and FF reference waveforms used in \eqref{eq:fbref} and \eqref{eq:FFref} may be different.

\subsubsection{Plasma Current Controller} \label{sec:plasma_controller}

The Plasma Current Controller regulates the total plasma current, ensuring that it follows the desired reference trajectory.
It does so by producing a request, of the same FF and FB structure as \eqref{eq:fb_ff}, for the rate of change of current in the Ohmic PF coil(s) (denoted here as $I_\Omega$).
The FB contribution is produced by a PID controller \eqref{eq:PID}, using the error
\begin{align*}
    \delta_{I_p}(t) = I_p^{\text{ref,FB}} (t) - I_p^{\text{meas}} (t)\ ,
\end{align*}
between the user-defined FB reference $I_p^{\text{ref,FB}}(t)$ and the measured plasma current $I_p^{\text{meas}}(t)$.
As before, the gains used by this controller, $(K_{I_p}^{\text{prop}}, K^{\text{int}}_{I_p}, K_{I_p}^{\text{der}})$, are user-defined waveforms.
The FF contribution enables the user to set a waveform for the loop voltage on the plasma ${V_{\text{loop}}^{\text{FF}}}(t)$. 
The combined FB and FF request is given by
\begin{align} \label{eq:plasma_alg}
    \left[\frac{\rd}{\rd t} I_\Omega\right]_{\text{req}} = b_{I_p} \left[\frac{\rd}{\rd t} I_p\right]_{\text{FB}}+ (1 - b_{I_p}) \frac{V_{\text{loop}}^{\text{FF}}}{\mathcal{M}_{{\Omega}}},
\end{align}
where ${\mathcal{M}_{\Omega}(t)}$ is a waveform for the mutual inductance between the plasma and the Ohmic coils and $b_{I_p}$ is the blend coefficient (recall \eqref{eq:fb_ff}).

\subsubsection{Shape Controller} \label{sec:shape_controller}

The Shape Controller regulates a set of user-specified shape parameters, that can include both core and divertor descriptors, such that they follow their specified reference trajectories. 
Such descriptors can take a variety of forms, including physical coordinates of specific plasma features (e.g. X-points, strikepoints, etc.); poloidal flux values at given locations; and closest ``gaps'' between, say, the plasma separatrix and a chosen point on the first wall.


The Shape Controller output is a request for the rate of change of each shape parameter, collected here in the vector $\bm{P}(t)$.
Following the same structure as \eqref{eq:fb_ff}, it includes both FB and FF contributions:
\begin{align} \label{eq:shape_alg}
    \left[\frac{\rd}{\rd t}\bm{P}\right]_{\text{req}} = \bm{b}_{\bm{P}} \left[\frac{\rd}{\rd t} \bm{P}\right]_{\text{FB}} + \left(1 -  \bm{b}_{\bm{P}} \right) \left[\frac{\rd}{\rd t} \bm{P}\right]_{\text{FF}},
\end{align}
where $\bm{b}_{\bm{P}}$ is a vector of blend coefficients. 
The FB contribution is produced by a PID controller \eqref{eq:PID} with error
\begin{align*}
    \bm{\delta}_{\bm{P}}(t) = \bm{P}^{\text{ref,FB}} (t) - \bm{P}^{\text{meas}} (t)\ ,
\end{align*}
and user-defined vectors of gains $(\bm{K}_{\bm{P}}^{\text{prop}}, \bm{K}^{\text{int}}_{\bm{P}},\bm{K}_{\bm{P}}^{\text{der}})$. 
The FB reference waveform is denoted by $\bm{P}^{\text{ref,FB}}(t)$ and the measurement of the shape parameters by $\bm{P}^{\text{meas}}(t)$.
The FF contribution is as in \eqref{eq:FFref} and uses a FF reference waveform $\bm{P}^{\text{ref,FF}} (t)$.

\subsubsection{Virtual Circuits Controller} \label{sec:vc_controller}

The Virtual Circuits (VCs) Controller combines the requests from the Plasma and Shape Controllers, transforming them into requests for the rate of change of currents in the PF coils using VCs. 

First, let us focus on control of the magnetic topology.
In order to move each shape parameter independently of all the others, a VC matrix can be used. 
A VC matrix $\mathcal{V}$ is defined as the pseudoinverse of the sensitivity matrix, which itself is the Jacobian of the shape parameters with respect to currents in the PF coils, for a specific equilibrium \citep[e.g.,][]{lvovskiy2025}. 
Each column of the VC matrix gives the changes in PF coil currents required to change a given shape parameter by one unit, without changing any of the other shape parameters considered (i.e. it decouples shape parameter changes from one another).
VC matrices are, however, a function of the equilibrium from which they were generated and so this decoupling becomes less accurate when a VC matrix is used to control a different equilibrium from the one it was derived for. 
The Virtual Circuits Controller therefore enables use of a schedule of VC matrices ${\mathcal{V}}(t)$ to be used between different times in the simulation. 

These VCs enable us to transform the request \eqref{eq:shape_alg} into a request for the rate of change of PF coil currents:
\begin{align} \label{shape_unapprov}
    \left[\frac{\rd}{\rd t} \bm{I} \right]_{\text{req},\bm{P}} = \mathcal{V}_{\bm{P}}(t) \left[\frac{\rd}{\rd t} \bm{P}\right]_{\text{req}}\ .
\end{align}
Note that $\mathcal{V}_{\bm{P}}(t)$ is the VC matrix specific to the shape parameters in $\bm{P}$, while below we will use $\bm{\mathcal{V}}_{\Omega}(t)$ to denote the VC for the Ohmic circuit.
We have also dropped the subscript from $\bm{I}_\text{act}$, used to denote the PF coil currents, for conciseness.

In a similar way, the rate of change of PF coil currents resulting from the request made by the Plasma Current Controller is
\begin{align} \label{ohmic_unapprov}
    \left[\frac{\rd}{\rd t} \bm{I}\right]_{\text{req},\Omega} = {\bm{\mathcal{V}}_\Omega}(t) \left[\frac{\rd}{\rd t} I_\Omega\right]_{\text{req}}\ .
\end{align}
Instead of being a matrix, $\bm{\mathcal{V}}_\Omega$ is a vector of dimensionless weights defining the Ohmic circuit, which is (ideally) built such that its contributions to the total poloidal flux do not cause the plasma shape parameters to change \citep[e.g.,][]{mcardle2020}.
If so, the request \eqref{ohmic_unapprov} should not cause undesired movements of the plasma parameters beyond those driven by \eqref{shape_unapprov}.

These requests are then combined with any user-defined FF coil current requests $\bm{I}^{\text{ref,FF}}(t)$, such that
\begin{align} \label{eq:unapp_currents_deriv}
    \left[\frac{\rd}{\rd t} \bm{I}\right]_{\text{req}} = \left[\frac{\rd}{\rd t} \bm{I}\right]_{\text{req},\bm{P}} +  \left[\frac{\rd}{\rd t} \bm{I}\right]_{\text{req},\Omega}\ + \frac{\rd}{\rd t} \bm{I}^{\text{ref,FF}}, 
\end{align}
giving the final output from the Virtual Circuits Controller.

\subsubsection{Systems Controller} \label{sec:systems_controller}

The Systems Controller enforces PF coil current limits for machine protection and safe tokamak operation to the requests from the Virtual Circuits Controller.
First, limits are applied to the requested rate of change of the PF coil currents:
\begin{align*}
    \left[\frac{\rd}{\rd t} \bm{I}\right]_{\text{req}}^{\text{appr}} = \min \left(|\dot{\bm{I}}_{\text{ramp}}|, \max\left(-|\dot{\bm{I}}_{\text{ramp}}| ,\left[\frac{\rd}{\rd t} \bm{I}\right]_{\text{req}}\right)\right),
\end{align*}
where $|\dot{\bm{I}}_{\text{ramp}}|(t)$ denotes the user-defined vector of maximum allowed rate of change per coil and $\left[{\rd}\bm{I}/{\rd t} \right]_{\text{req}}^{\text{appr}}$ is the resulting approved request. 
These are then integrated to obtain a set of requested PF coil currents:
\begin{align*}
    \left[ \bm{I}(t) \right]_{\text{req}} = \left[\bm{I}(t-\rd t)\right]_{\text{req}}^{\text{appr}} + \left[\frac{\rd}{\rd t} \bm{I}\right]_{\text{req}}^{\text{appr}} \rd t.
\end{align*}
These are then optionally clipped based on user-defined upper $\bm{I}^{\text{up}}(t)$ and lower $\bm{I}^{\text{low}}(t)$ coil current limits, to obtain the final approved current requests
\begin{align} \label{eq:approved_current_clip}
    \left[\bm{I}(t)\right]_{\text{req}}^{\text{appr}} = \min \left( \bm{I}^{\text{up}},\, \max\left( \bm{I}^{\text{low}},\, \left[\bm{I}(t)\right]_{\text{req}} \right) \right).
\end{align}


\subsubsection{PF Voltages Controller} \label{sec:pf_controller}

The PF Voltages Controller transforms the approved coil current requests from the Systems Controller into the final voltage requests. 

The first contribution to the voltage requests is obtained using a set of simplified circuit equations (CE):
\begin{align}\label{eq:CE_req}
    \left[\bm{V} (t)\right]_{\text{req,CE}} = {\mathcal{R}}(t) \bm{I}^{\text{meas}} (t) + {\mathcal{M}}^{\text{FF}}(t) \left[\frac{\rd}{\rd t} \bm{I}\right]_{\text{req}}^{\text{appr}}.
\end{align}
Here, ${\mathcal{R}}(t)$ and ${\mathcal{M}}^{\text{FF}}(t)$ are user-defined resistance and mutual inductance matrices for the PF coils. 
With respect to the circuit equations described in \cref{sec:evolution}, those in \eqref{eq:CE_req} are substantially simplified and do not account for the plasma or the passive  structures.
To compensate for this, the matrices ${\mathcal{R}}(t)$ and ${\mathcal{M}}^{\text{FF}}(t)$ will be different to those defined in \cref{sec:evolution} and can be defined as functions of time if required.

The second contribution is FB driven and is of the form
\begin{align}\label{eq:PF_FB}
    \left[\bm{V} (t)\right]_{\text{req,FB}} = \mathcal{M}^{\text{FB}}(t) \left[\frac{\rd}{\rd t} \bm{I}\right]_{\text{FB}},
\end{align}
where a second user-defined mutual inductance matrix, ${\mathcal{M}}^{\text{FB}}(t)$, potentially different from the one used in \eqref{eq:CE_req}, can be adopted. 
The FB derivative on the right hand side is produced by a dedicated PID controller with gains $(\bm{K}_{\bm{I}}^{\text{prop}}, \bm{K}^{\text{int}}_{\bm{I}}, \bm{K}_{\bm{I}}^{\text{der}})$ and
based on the discrepancy
\begin{align*}
    \bm{\delta}_{\bm{I}}(t) = \left[\bm{I}(t)\right]_{\text{req}}^{\text{appr}} - \bm{I}^{\text{meas}} (t),
\end{align*}
between the approved active coil current requests and the measured PF coil currents $\bm{I}^{\text{meas}} (t)$. 
Combining \eqref{eq:CE_req} and \eqref{eq:PF_FB}, we get the total (unapproved) voltage requests
\begin{align}
    \left[\bm{V} (t)\right]_{\text{req}} = \left[\bm{V} (t)\right]_{\text{req,CE}} + \left[\bm{V} (t)\right]_{\text{req,FB}}\ .
\end{align}

These requests are also clipped to adhere to any operational limits.
First, a clip can be applied to the voltage requests themselves:
\begin{align*}
     \left[\bm{V} (t)\right]_{\text{req}}^{\text{clip}} = \min \left(  |\bm{V}_{\text{lim}}|, \max \left( -|\bm{V}_{\text{lim}}|,  \left[\bm{V} (t)\right]_{\text{req}}  \right) \right),
\end{align*}
where the limits $\bm{V}_{\text{lim}}(t)$ are user-defined waveforms.
Then, a clip can be applied to the derivatives of the voltage requests using $|\dot{\bm{V}}_{\text{ramp}}|(t)$, and built as follows
\begin{align*}
 \left[\frac{\rd}{\rd t} \bm{V}\right]_{\text{req}} = \left(\left[\bm{V} (t)\right]_{\text{req}}^{\text{clip}}- \left[\bm{V} (t-\rd t)\right]_{\text{req}}^{\text{appr}}\right)/\rd t, 
\end{align*}
so that
\begin{align*}
     \left[\frac{\rd}{\rd t} \bm{V}\right]_{\text{req}}^{\text{appr}} = \min \left(  |\dot{\bm{V}}_{\text{ramp}}|, \max \left( -|\dot{\bm{V}}_{\text{ramp}}|,  \left[\frac{\rd}{\rd t} \bm{V}\right]_{\text{req}} \right) \right).
\end{align*}
With this, we can build the approved voltage requests
\begin{align}
 \left[\bm{V} (t)\right]_{\text{req}}^{\text{appr}} = \left[\bm{V} (t-\rd t)\right]_{\text{req}}^{\text{appr}}  + \left[\frac{\rd}{\rd t} \bm{V}\right]_{\text{req}}^{\text{appr}} \rd t, 
\end{align}
which are passed to the simulator along with the voltage to be applied to the vertical control coil \eqref{eq:vert_req} and coil resistances \eqref{eq:coil_activations}. 

\subsubsection{Vertical Controller} \label{sec:vertical_controller}

Elongated plasmas are inherently vertically unstable \citep[e.g.,][Chp. 4.6.6]{jardin2010} and require active feedback control. 
The PF coils used for vertical control on MAST-U are separate from the plasma shaping coils and consist of an up/down symmetric pair of coils wired in anti-series.
These generate a radial magnetic field, and hence up/down forces that stabilise the plasma vertically.

The Vertical Controller in the FPDT is a purely FB driven controller, based on the error
\begin{align*}
    \delta_{zI_p}(t) = I_p^{\text{meas}}(t) \left( z^{\text{ref,FB}} (t) - z^{\text{meas}}(t) \right),
\end{align*}
where $z^{\text{ref,FB}}(t)$ is the user-defined reference waveform and $I_p^{\text{meas}}(t)$ the measured plasma current. 
We measure the average vertical coordinate of the plasma current distribution using
\begin{align} \label{eq:vert_pos}
    z^{\text{meas}}(t) = {\frac{1}{I_p^{\text{meas}}(t)}} \int_{\Omega_p} Z J_{\phi}(\psi, R, Z; \bm{\theta}) \ \mathrm{d}R \mathrm{d}Z,
\end{align}
noting that $J_{\phi}$ is measured at time $t$, though not explicitly defined here. 
A dedicated PID controller uses this discrepancy to produce the following voltage request
\begin{align}\label{eq:vert_req}
    V_{zI_p}(t) = K_{zI_p}^{\text{prop}} \delta_{zI_p} + K^{\text{int}}_{zI_p}\iota_{zI_p} +K_{zI_p}^{\text{der}}\frac{\rd}{\rd t}\delta_{zI_p},
\end{align}
which is applied to the vertical control coils.
The gains used by this controller, $(K_{zI_p}^{\text{prop}}, K^{\text{int}}_{zI_p}, K_{zI_p}^{\text{der}})$, are provided as user-defined waveforms.

\subsubsection{Coil Activation Controller} \label{sec:coil_act_controller}

The Coil Activation Controller handles the timings at which the PF coils are switched on during the shot---an essential feature for modelling the pre-plasma and plasma ramp-up phases when coils can only be activated in certain sequences.
Rather than enforcing this by explicitly varying the number of active coils at different times, which would require changes to the machine description itself, we do so instead by altering the resistance matrix $\mathcal{R}_{m,m}$ dynamically, recall \eqref{eq:circuit_eqs}.
A PF coil can effectively be disconnected (switched off) by setting its resistance to an arbitrarily large value, denoted here as $\Omega_\infty$.

In \eqref{eq:circuit_eqs}, we therefore overwrite the default $\mathcal{R}_{m,m}$ with
\begin{align} \label{eq:coil_activations}
    \mathcal{R}_{m,m}^{\text{ctrl}} (t) = {\mathcal{R}}_{m,m} + \Omega_\infty  \text{diag} \left( \bm{1} - \bm{h}^{\text{activation}} (t) \right),
\end{align}
where $\bm{h}^{\text{activation}}(t)$ is a vector of boolean values indicating the on $(1)$ and off $(0)$ state of each coil at time $t$ and $\bm{1}$ is a vector of ones.

While the ability to modify resistances during a simulation is \textit{not} something a real PCS would do, this functionality exploits the fact that the virtual PCS and simulator coexist as objects within the same computational framework.
This enables us to not only switch coils on or off, but also test deliberate mismatches between the controller's assumed parameters and those of the physical model---for example in the resistance matrices as a means of probing robustness to calibration errors or sensitivity to model uncertainty.

\section{Validation using MAST-U discharges} \label{sec:results}

\subsection{Overview}

\subsubsection{Objectives}

Before using the FPDT for predictive simulations, it is essential we validate its performance by comparing to experimental data. 
We do so by re-simulating past MAST-U shots: the close proximity of the MAST-U PF coils to the plasma and each other drive strong cross-coupling effects, making plasma control particularly challenging and therefore an excellent test case for the FPDT. 

On one hand, we aim to demonstrate that the FPDT can effectively sustain and control the evolution of plasma current, vertical position, and key shape parameters, as requested by the reference waveforms.
In order to do so, we will quantify any discrepancy between the evolution of the FB controlled parameters $X(t)$ (whether from simulation or experiment) with the desired FB reference waveform using the mean absolute difference (MAD):
\begin{align}\label{MAD}
    \text{MAD}(X) = \frac{1}{N+1} \sum_{k=0}^{N} \left| X(t_k) - X^{\text{ref,FB}}(t_k) \right| ,
\end{align}
and the relative mean absolute difference (RMAD):
\begin{align}\label{RMAD}
    \text{RMAD}(X) = \frac{1}{N+1} \sum_{k=0}^{N} \frac{\left| X(t_k) - X^{\text{ref,FB}}(t_k) \right|}{\left| X^{\text{ref,FB}}(t_k) \right|},
\end{align}
where $N$ denotes the number of time steps in the simulation/experiment. 

On the other hand, we wish to compare simulation results to the experimental measurements themselves.
As we provide the FPDT with the same control settings the actual MAST-U PCS was provided with, we should find agreement between our simulations and the experimental data. 
However, it is worth highlighting that this latter comparison has clear limitations: 
\begin{itemize}
    \item The experimental data is affected by other forms of control (e.g. fuelling, impurity seeding, etc.), and by transient events (e.g. internal reconnection events, edge localised modes) that are not captured by the FPDT simulations.
    \item The FDPT extracts measurements of controlled quantities from the simulated equilibria, whereas experimental values are obtained from magnetic measurements via LEMUR \citep{kochan2023}, which can be subject to noise. 
    Any discrepancies and biases introduced by this difference are left for future analysis.
\end{itemize}

\subsubsection{Simulation setup}
The MAST-U machine description includes thirteen active PF coils and $150$ passive structure elements (see \cite{pentland2024}).
For the PF coils, we calculate calibrated resistances and mutual inductances using vacuum shot data, whereas for the passive structures, we use purely geometric values. 
All presented simulations are based on a reduction of the passive structures to $30$ eigenmodes---those that couple most strongly to the plasma. 
We use a $65 \times 65$ computational grid and a fixed time step of \SI{0.2}{\milli\second}, small enough to resolve the vertical dynamics of the plasmas we encounter. 
We do, however, run the virtual PCS at twice this frequency, i.e. every \SI{0.1}{\milli\second}, so that we match the frequency that the actual MAST-U PCS runs at.
An update of the linearisation used by the PwLD simulation mode is triggered when the relative change in plasma current density exceeds \SI{5}{\percent}.
The update is triggered in the PwL mode when any of the core shape parameters (see next section) change by more than \SI{1}{\centi\meter}; the radial nose position by \SI{5}{\centi\meter}; the strike point\footnote{This criteria is larger because the linearisation does not lose significant accuracy when doing a leg sweep---it primarily uses the outer divertor coils, which do not impact the other shape parameters significantly.} by \SI{15}{\centi\meter}; or the vertical position by \SI{2.5}{\milli\meter}.
In addition, for the NL mode, the relative convergence tolerances for the dynamics \eqref{eq:circuit_eqs}-\eqref{eq:plasma_eq} and GS \eqref{eq:GS_PDE} are set to \SI{1}{\percent}, i.e. the residuals in the extensive currents and plasma flux are required to be smaller than \SI{1}{\percent} of their respective change in each time step (we refer the reader to \citet[eqs.~25 and~26]{amorisco2024} for further details).

\begin{figure}[t]
    \begin{subfigure}{0.99\linewidth}
        \centering
        \includegraphics[width=0.8\textwidth]{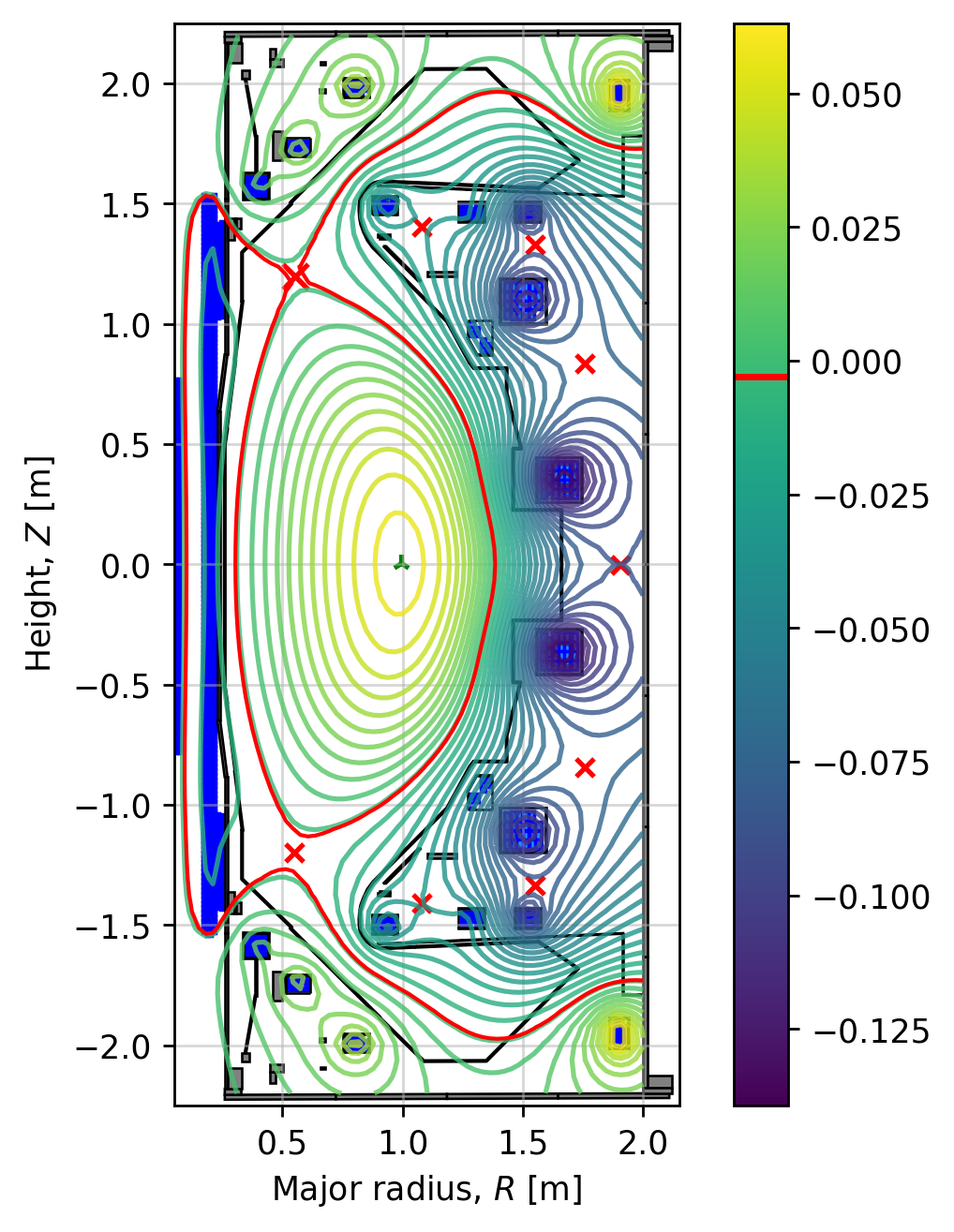}
    \end{subfigure}
    \caption{
    FPDT-simulated equilibrium of MAST-U shot $52570$ ($t = 0.55s$) with $\psi$ contours (see colour bar) in $\Omega = [0.06,2] \times [-2.2,2.2]$. 
    Key elements include the separatrix (red line), X-points (red crosses), magnetic axis (green marker), and the primary X-point (larger red cross). 
    Also shown are the thirteen active poloidal field coils (dark blue), the passive structures (dark grey), and the wall/limiter (solid black).
    }
    \label{fig:52570_eq}
\end{figure}

The simulations here are focused on the flat-top phase, $t \geq \SI{0.2}{\second}$, with future work aiming to extend this to the ramp-up phase.
We start our simulations with static equilibria built using EFIT\texttt{++} reconstruction data \citep{kogan2022}, including plasma, coil, and passive structure currents, as well as the associated plasma profiles (see below).
Through testing, we found that it is often useful to partially nudge these initial equilibria towards the reference waveforms before simulation as the intrinsic uncertainties in the EFIT\texttt{++} reconstructions introduce exceedingly large initial departures, affecting the subsequent time evolution in a spurious manner.

In the absence of self-consistent plasma profile evolution, we prescribe the parameters $\bm{\theta}(t)$ as a function of time by using the magnetics-only EFIT\texttt{++} reconstruction data for the given shot. 
This prescribed evolution takes into account the heating and current drive generated by the neutral beam injectors on MAST-U, as FreeGSNKE does not currently account for any non-inductive current drive (whose contribution on MAST-U is small anyway, see \cite{turnyanskiy2009}).
Similarly, the time trace of the plasma resistivity $\rho_p(t)$ is also prescribed. 
We obtain an estimate of the plasma resistivity in the following MAST-U shots from separate ``open-loop'' simulations. 
Instead of using our virtual PCS class, in these we employ the voltages recorded by the the real MAST-U PCS (with the exception of the vertical control coil which we continue to control in feedback with the virtual PCS). 
The resistivity of the plasma is then tuned by a custom controller so to that the evolution of the total plasma current is matched to the FB waveform. 
This offers a rapid approximation of the plasma resistivity for use in the ``closed-loop'' FPDT simulations that follow.

\subsection{Shot 52570 (Super-X divertor)} \label{sec:super-X}

\begin{figure}[t!]
    \begin{subfigure}{0.99\linewidth}
        \centering
        \includegraphics[width=0.99\textwidth]{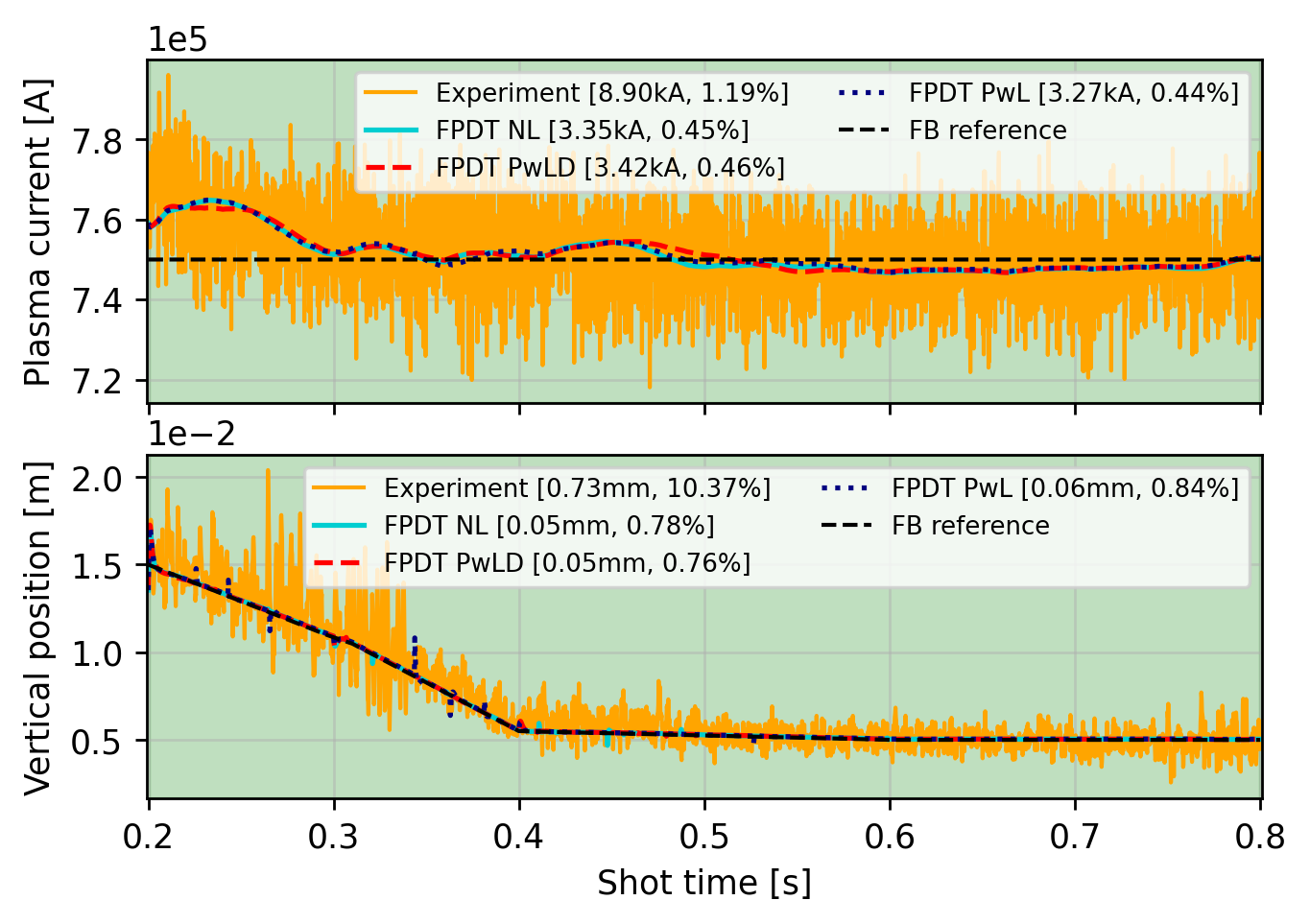}
    \end{subfigure}
    \caption{
    FPDT-simulated evolution of total plasma current (top panel) and vertical position (bottom panel) for MAST-U shot $52570$ using the NL (solid light blue), PwLD (dashed red), and PwL (dotted dark blue) simulation modes.
    Also shown are the FB reference waveforms used in the real and virtual PCS (dashed black) and the measured experimental data from the discharge (orange). 
    Background shading indicates when FB control is on (green), when FF control is on (yellow), and when no control occurs (white)---see later figures for other shading. 
    }
    \label{fig:52570_ip}
\end{figure}

\begin{figure}[t!]
    \begin{subfigure}{0.99\linewidth}
        \centering
        \includegraphics[width=0.99\textwidth]{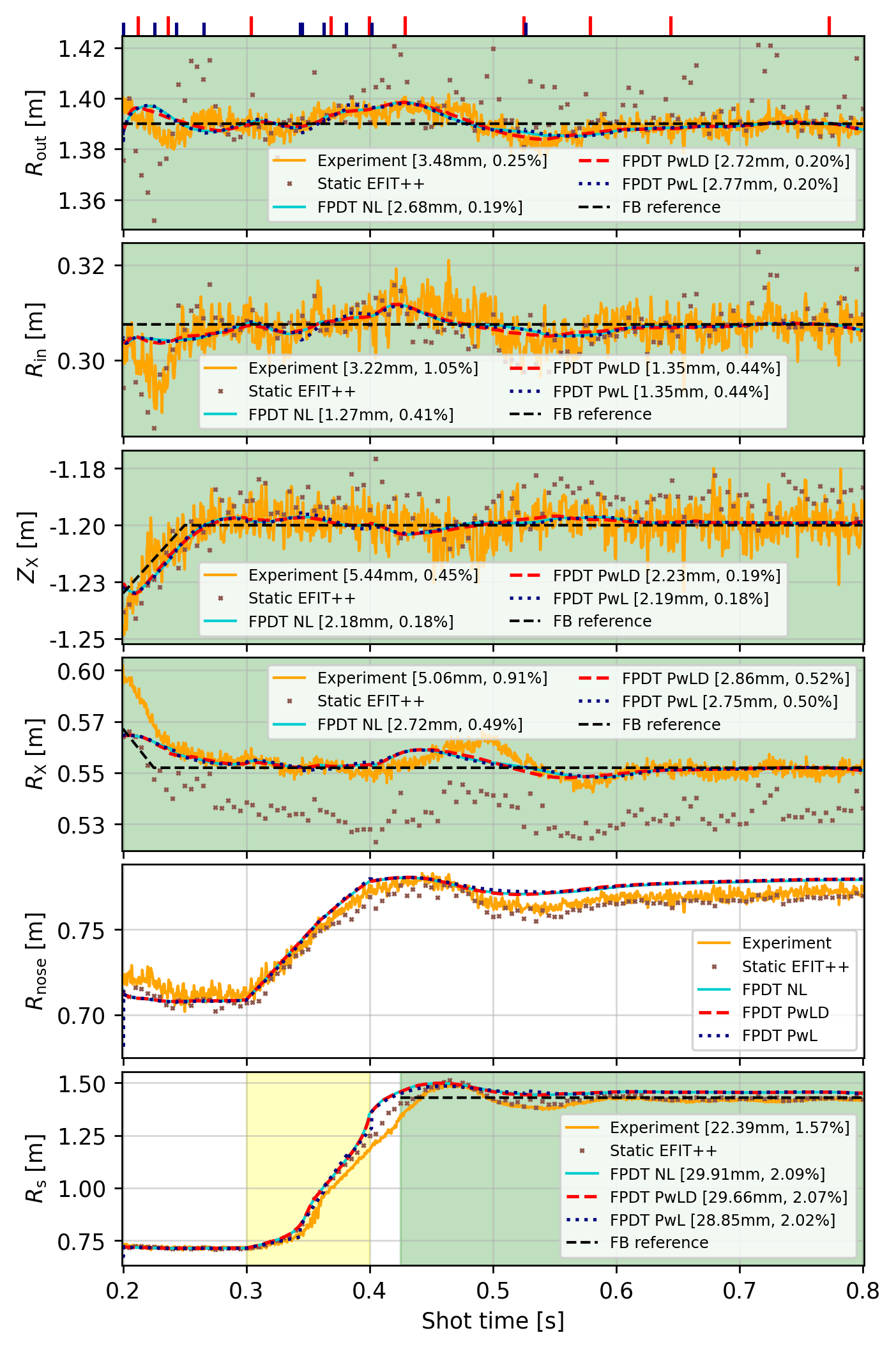}
    \end{subfigure}
    \caption{
    FPDT-simulated evolution of several shape parameters (as described in the main text) for MAST-U shot $52570$ using the NL (solid light blue), PwLD (dashed red), and PwL (dotted dark blue) simulation modes.
    Also shown are the FB (dashed black) reference waveforms used in the real and virtual PCS, alongside the LEMUR measured experimental data from the discharge (orange) and values from static FreeGSNKE-simulated equilibria using EFIT\texttt{++} reconstruction data (brown crosses).
    Background shading indicates when FB control is on (green), when FF control is on (yellow), and when no control occurs (white). 
    Tick marks above the top panel indicate times at which relinearisation took place in the PwLD (red) and PwL (dark blue) simulations. 
    }
    \label{fig:52570_shapes}
\end{figure}

Here, we re-simulate the flat-top phase of MAST-U shot $52570$, a double null \SI{750}{\kilo\ampere} plasma heated by two sets of neutral beams, designed to demonstrate the formation of a super-X divertor \citep{valanju2009, morris2014,morris2018}.
This is a key configuration for MAST-U divertor studies, as it sweeps the divertor strikepoints to a large major radius, thereby enabling larger connection lengths, flux expansion, and area over which heat flux is deposited on divertor tiles (see \cref{fig:52570_eq}).
ll of the following time series results display real-time LEMUR measurements available to the MAST-U PCS in orange, together with results from the different FPDT simulation modes. 
The legends display values of the metrics \eqref{MAD} (with units) and \eqref{RMAD} (with percentage), comparing the FB reference waveforms to both simulation and experimental measurements.
Green background shading represents when FB control is on, yellow when FF control is on, and white when no control is used.

In \cref{fig:52570_ip}, we focus on plasma current (top panel) and vertical position (bottom panel). Excellent agreement can be seen in both, with all simulation modes achieving RMAD values below \SI{0.5}{\percent} for the plasma current and \SI{1}{\percent} for the vertical position. 

In \cref{fig:52570_shapes}, we plot the evolution of six commonly controlled shape parameters in MAST-U \citep{anand2024}, namely:
\begin{itemize}
    \item $R_{\text{in}}$ and $R_{\text{out}}$: the inboard and outboard midplane radii. 
    \item $R_{\text{X}}$ and $Z_{\text{X}}$: the radial and vertical positions of the X-point near the lower divertor chamber. 
    \item $R_{\text{nose}}$: the radial intersection between the lower divertor leg (separatrix) and a virtual line between the centre of the DP and D2 coils---see \citet[Fig. 1]{pentland2024}.
    \item $R_{\text{s}}$: the radial intersection between the lower divertor leg and the divertor chamber tiles.
\end{itemize}
In addition to the real-time LEMUR data, each panel of \cref{fig:52570_shapes} also displays values of the same shape parameter obtained from EFIT\texttt{++} reconstruction data (brown crosses).
The size of fast variations in the EFIT\texttt{}{++} data and the magnitude of the discrepancy sometimes present between LEMUR and EFIT\texttt{++} measurements of the same shape parameter provide a qualitative scale for the typical uncertainty in these quantities. 
In turn, this scale is useful when contextualising the size of any discrepancies between simulations and experimental results.
In the first two panels, we can see that both $R_{\text{in}}$ and $R_{\text{out}}$ are successfully held at their respective constant FB reference values throughout the simulations in all cases, with MAD values no larger than those from the experimental results.
Correspondingly, the third panel shows $Z_{\text{X}}$ is successfully shifted upwards over \SI{3}{\centi\meter} from its starting value in little under \SI{50}{\milli\second} and, again, held constant thereafter (with RMAD values less than \SI{0.25}{\percent}).
Similarly, $R_{\text{X}}$ is successfully moved inwards in the fourth panel according to the FB waveform (albeit with slightly larger RMAD values than $Z_{\text{X}}$, reflected in the uncertainty with the EFIT\texttt{++} values).
The fifth panel displays the uncontrolled parameter $R_{\text{nose}}$ evolving similarly to both experimental LEMUR and EFIT\texttt{++} values.
In the final panel, the strike point is swept outwards using a FF drive between \SI{300}{\milli\second} and \SI{400}{\milli\second}, moving from a conventional divertor scenario to a super-X one. 
After a short period of no control, the super-X divertor is thereafter maintained successfully in FB.

\begin{figure*}[t!]
    \begin{subfigure}{0.99\linewidth}
        \centering
        \includegraphics[width=0.99\textwidth]{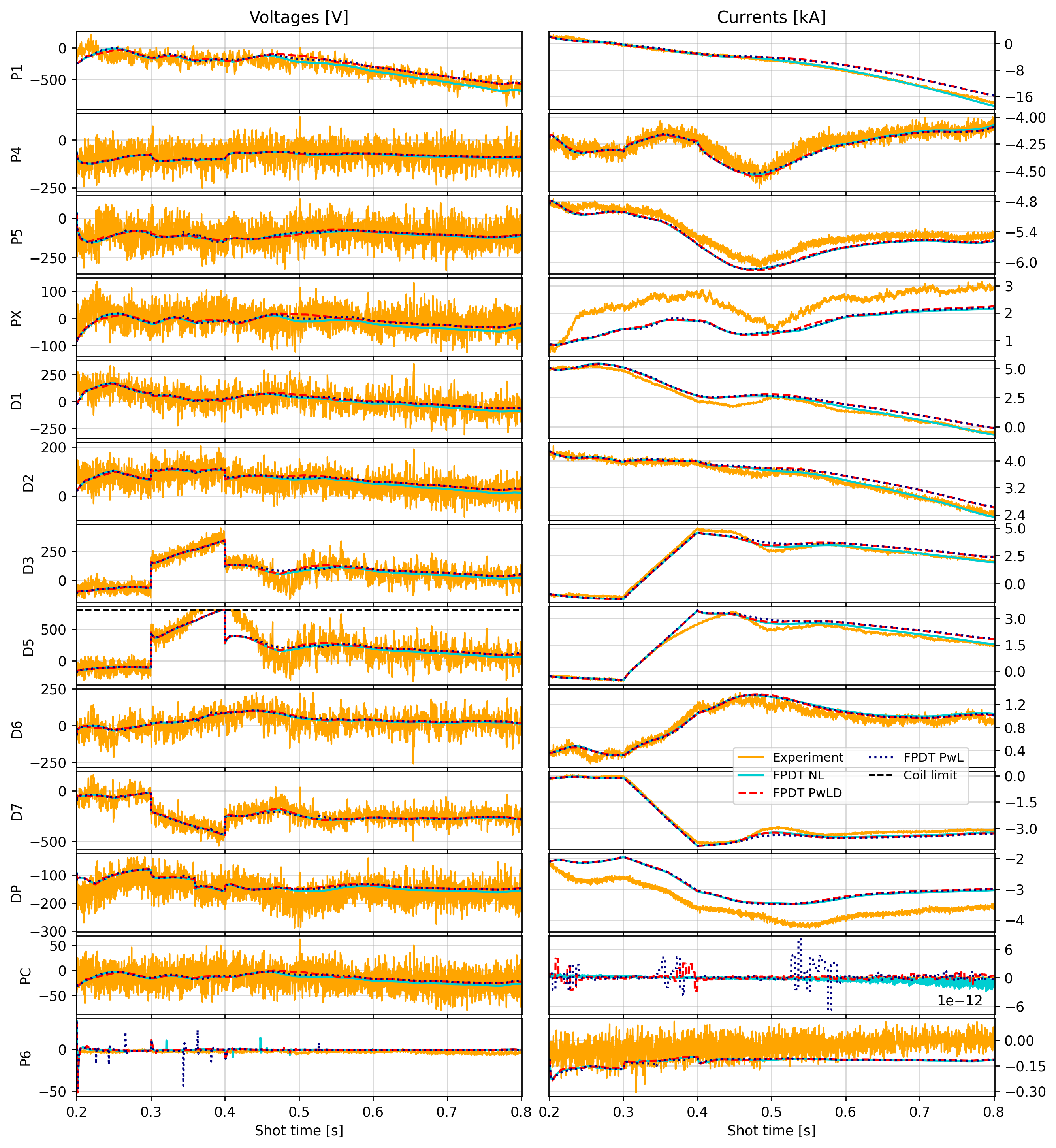}
    \end{subfigure}
    \caption{
    FPDT-simulated evolution of voltages (left panels) and currents (right panels) in all thirteen active PF coils (top to bottom) for MAST-U shot $52570$ using the NL (solid light blue), PwLD (dashed red), and PwL (dotted dark blue) simulation modes.
    Also shown in some panels are the coil voltage and current limits (dashed black) alongside the measured experimental values from the discharge (orange).
    }
    \label{fig:52570_currents}
\end{figure*}

\cref{fig:52570_currents} displays the evolution of the PF coil voltages produced by the virtual PCS (left panels) and the subsequent evolution of the currents (right panels) in the FPDT, compared to the experimentally recorded values. 
Although not shown here, the passive structure currents are also evolved and can be monitored. 
We observe excellent overall qualitative agreement between the calculated voltages in all three simulation modes, noting in particular the correct application of the voltage limits---which are briefly reached on the D5 coil.
During the first \SI{10}{\milli\second} of simulation, however, larger than normal differences can be seen, driven primarily by difference between the actual and simulated initial equilibrium conditions, which we inherit from the EFIT++ initial conditions.
We can clearly see the impact that the FF phase on $R_{\text{S}}$ between \SI{300}{\milli\second} and \SI{400}{\milli\second} has on the divertor (D) coils with sharp jumps in voltages captured well by the virtual PCS.
We should also note that the PC coil is disconnected from the machine, but included in our machine model: a voltage is calculated by the PCS but no current is generated.
We should note that for the P6 voltage, in reality, $z_{meas}$ is not actually monitored in real-time on MAST-U; instead, the change in poloidal flux at two locations either side of the midplane is used as a proxy for vertical position.
While both measurements serve the same purpose, the P6 voltages produced by the FPDT vertical controller here will be slightly different from those in actual MAST-U discharges.

Despite starting very close to the experimental values (and with the correct voltages applied), a few of the coil currents (PX and DP) diverge after a short period of time, even though, as we saw, the plasma current, shape, and position all evolve correctly.
This may suggest that the calibration of our machine model (or perhaps unmodelled 3D effects), including the coil resistances and mutual inductances, can potentially be improved.
This said, the remaining currents match well; in particular the solenoid (P1), indicating that the resistivity has been well estimated.


All simulations reported were run on a single workstation (MacBook Pro, Apple M2 Pro chip, 16 GB RAM); no specialised or parallel hardware was used.
The total runtime for the NL simulation mode was $81\,$min \SI{11}{\second}, clearly rendering it prohibitively expensive for inter-shot scenario development and control design.
It does, however, provide us with a useful baseline to assess the accuracy of the PwLD and PwL simulation modes, both of which performed comparably to the NL with runtimes an order of magnitude smaller. 
The PwLD ran in $4\,$min \SI{32}{\second}, updating its linearisation $10$ times, while the PwL ran in $1\,$min \SI{51}{\second}, also relinearising $10$ times.
The times at which linearisation updates took place are indicated above the first panel in \cref{fig:52570_shapes}. 
As might be expected, most relinearisation takes place when the plasma shape changes significantly, i.e. early in simulations or when strike point sweeping takes place.
Despite improved performance in the PwLD and PwL modes, the dominant computational bottleneck remains the cost of relinearising around the evolving equilibrium, together with uncertainty in how frequent this is required.
There are, however, ways to mitigate this by using fewer passive structure modes or possibly re-using existing linearisations when those are applicable.


\begin{figure}[t!]
    \begin{subfigure}{0.99\linewidth}
        \centering
        \includegraphics[width=0.8\textwidth]{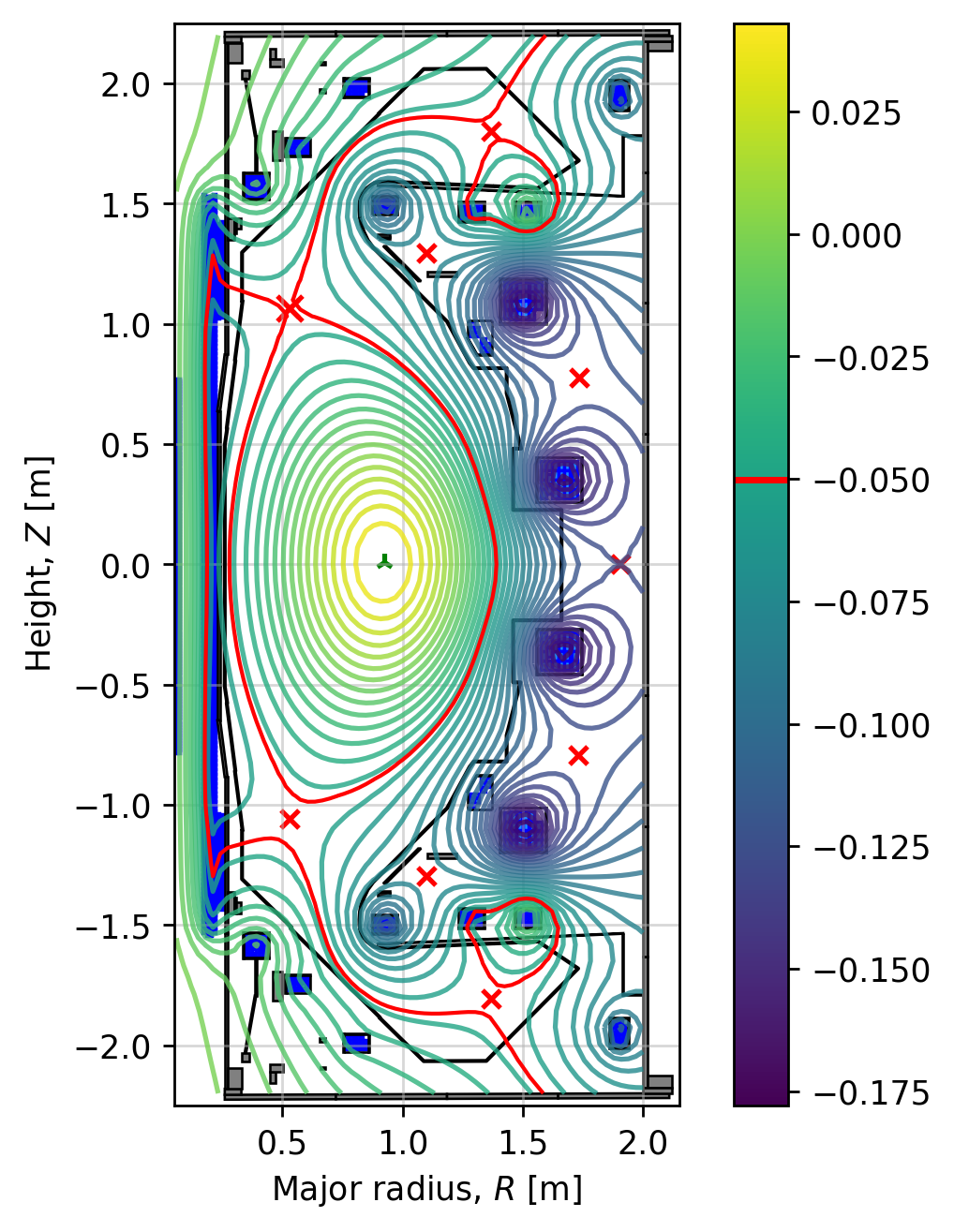}
    \end{subfigure}
    \caption{
    FPDT-simulated equilibrium of MAST-U shot $50366$ ($t = 0.50s$) with $\psi$ contours (see colour bar) in $\Omega = [0.06,2] \times [-2.2,2.2]$. 
    Key elements include the separatrix (red line), X-points (red crosses), magnetic axis (green marker), and the primary X-point (larger red cross). 
    Also shown are the thirteen active poloidal field coils (dark blue), the passive structures (dark grey), and the wall (solid black).
    }
    \label{fig:50366_eq}
\end{figure}
\begin{figure}[b!]
    \begin{subfigure}{0.99\linewidth}
        \centering
        \includegraphics[width=0.99\textwidth]{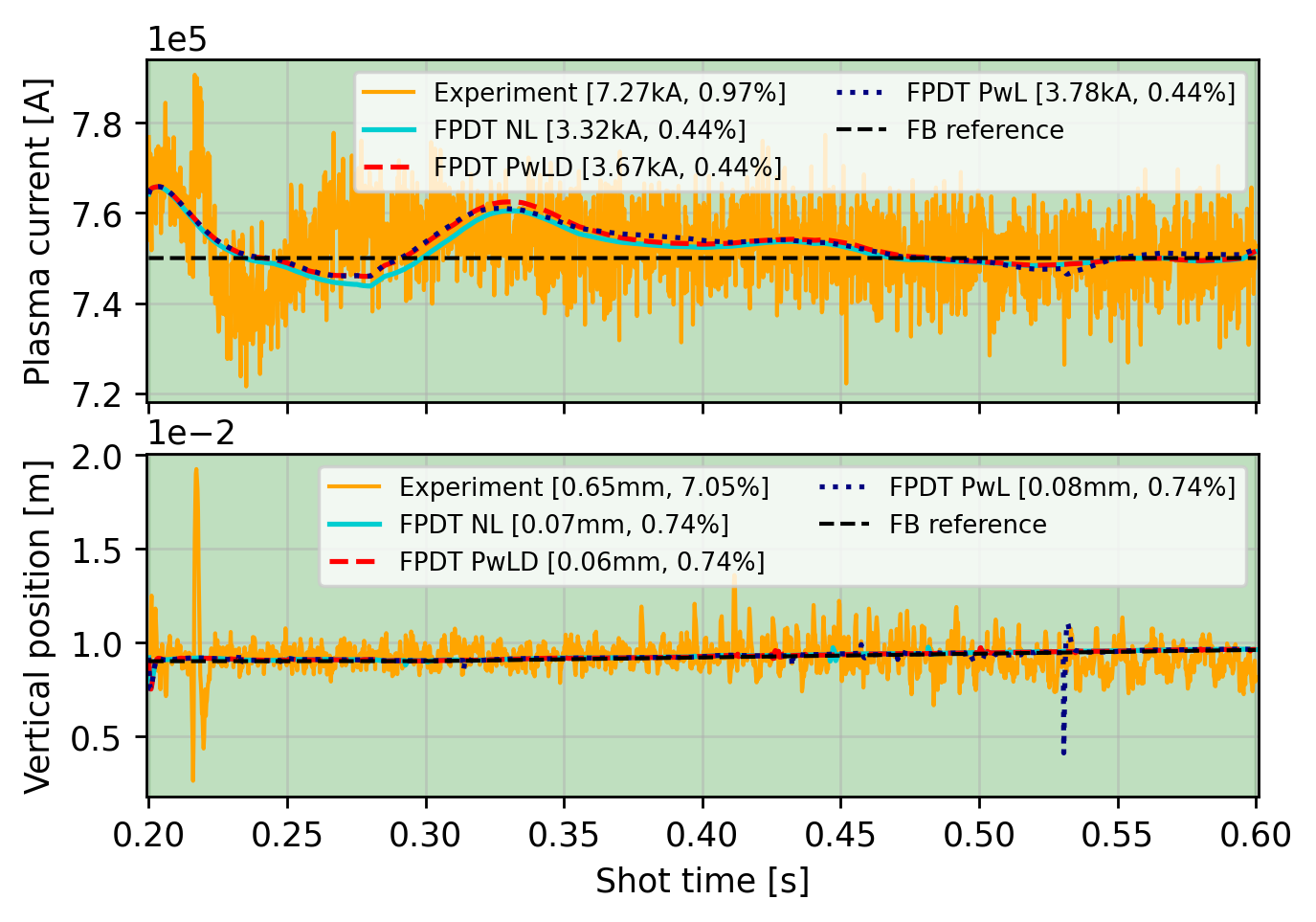}
    \end{subfigure}
    \caption{
    FPDT-simulated evolution of total plasma current (top panel) and vertical position (bottom panel) for MAST-U shot $50366$ using the NL (solid light blue), PwLD (dashed red), and PwL (dotted dark blue) simulation modes.
    Also shown are the FB reference waveforms used in the real and virtual PCS (dashed black) and the measured experimental data from the discharge (orange).
    See the main text for a description of the background shading convention.
    }
    \label{fig:50366_ip}
\end{figure}

\subsection{Shot 50366 (X-point target divertor)} \label{sec:x-point}

Next, we re-simulate the flat-top phase of MAST-U shot $50366$, a double null, \SI{750}{\kilo\ampere}, Ohmically heated plasma, which was designed to test the TED framework \citep[Tokamak Exhaust Designer, see][]{bardsley2025} and generate an X-point target divertor configuration.
The X-point target configuration generates a secondary X-point in the divertor chamber, close to the divertor leg (see \cref{fig:50366_eq}), in order to improve detachment access and reduce particle and heat fluxes \citep{labombard2015, lonigro2026}.

Similarly to shot $52570$, the simulated plasma current remains within \SI{0.5}{\percent} of the FB reference, despite being initialised over \SI{10}{\kilo\ampere} from the reference itself---see top panel of \cref{fig:50366_ip}.
The lower panel, again, demonstrates good vertical position control in all cases, tracking the steadily increasing FB reference. 

\begin{figure}[t!]
    \begin{subfigure}{0.99\linewidth}
        \centering
        \includegraphics[width=0.99\textwidth]{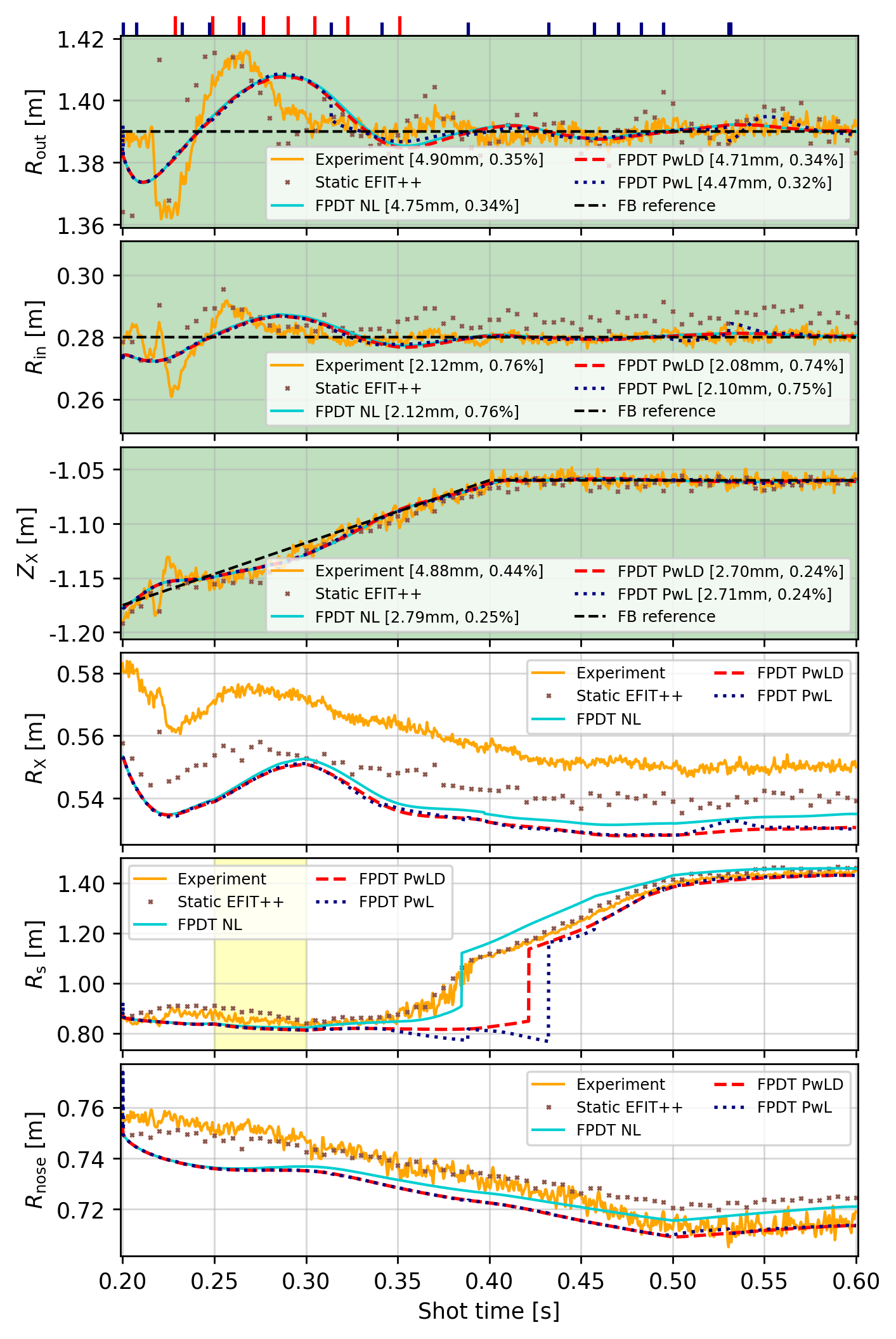}
    \end{subfigure}
    \caption{
    FPDT-simulated evolution of several shape parameters (as described in the main text) for MAST-U shot $50366$ using the NL (solid light blue), PwLD (dashed red), and PwL (dotted dark blue) simulation modes.
    Also shown are the FB (dashed black) reference waveforms used in the real and virtual PCS, alongside the LEMUR measured experimental data from the discharge (orange) and values from static FreeGSNKE-simulated equilibria using EFIT\texttt{++} reconstruction data (brown crosses).
    See the main text for a description of the background shading convention.
    Tick marks above the top panel indicate times at which relinearisation took place in the PwLD (red) and PwL (dark blue) simulations.
    }
    \label{fig:50366_shapes}
\end{figure}

As displayed in \cref{fig:50366_shapes}, the evolution of the shape parameters appears affected by the overshoot in $I_p$ in \cref{fig:50366_ip}, especially $R_{\text{out}}$ and $R_{\text{in}}$. 
Analogous initial oscillations are seen in both the LEMUR and EFIT\texttt{++} measurements. 
In our simulations they are successfully controlled to less than \SI{5}{\milli\meter} on average over the simulation, with peaks of approximately \SI{2}{\centi\meter} for $R_{\text{out}}$.
The ramp up in $Z_{\text{X}}$ (third panel) to form a smaller plasma core is tracked well, with the evolution of $R_{\text{X}}$ following the EFIT\texttt{++} reconstruction despite it being uncontrolled (fourth panel).
In this panel, and as previously noted \citep[Fig. 1(b)]{anand2024}, we observe a bias between the LEMUR measurements and EFIT\texttt{++} for $R_{\text{X}}$.
The bottom panel shows good qualitative agreement for $R_{\text{nose}}$, which is also uncontrolled. 

The fifth panel illustrates the evolution of $R_{\text{s}}$, which is not directly controlled besides a \SI{50}{\milli\second} period where $R_{\text{s}}$ is driven in FF. 
Each simulation mode is characterised by what appears to be a discontinuous strike point location, appearing sometime between \SI{350}{\milli\second} and \SI{450}{\milli\second}. 
\cref{fig:50366_strikepoint} illustrates the dynamics of the strike point in these simulations, which is in fact continuous: the apparent jumps are a result of the divertor leg `brushing against' the divertor tiles, running almost tangent to them. This phenomenology is not captured by the LEMUR data or by the EFIT\texttt{++} reconstructions, but is ubiquitous in our simulation modes.
\begin{figure}[t!]
    \begin{subfigure}{0.99\linewidth}
        \centering
        \includegraphics[width=0.99\textwidth]{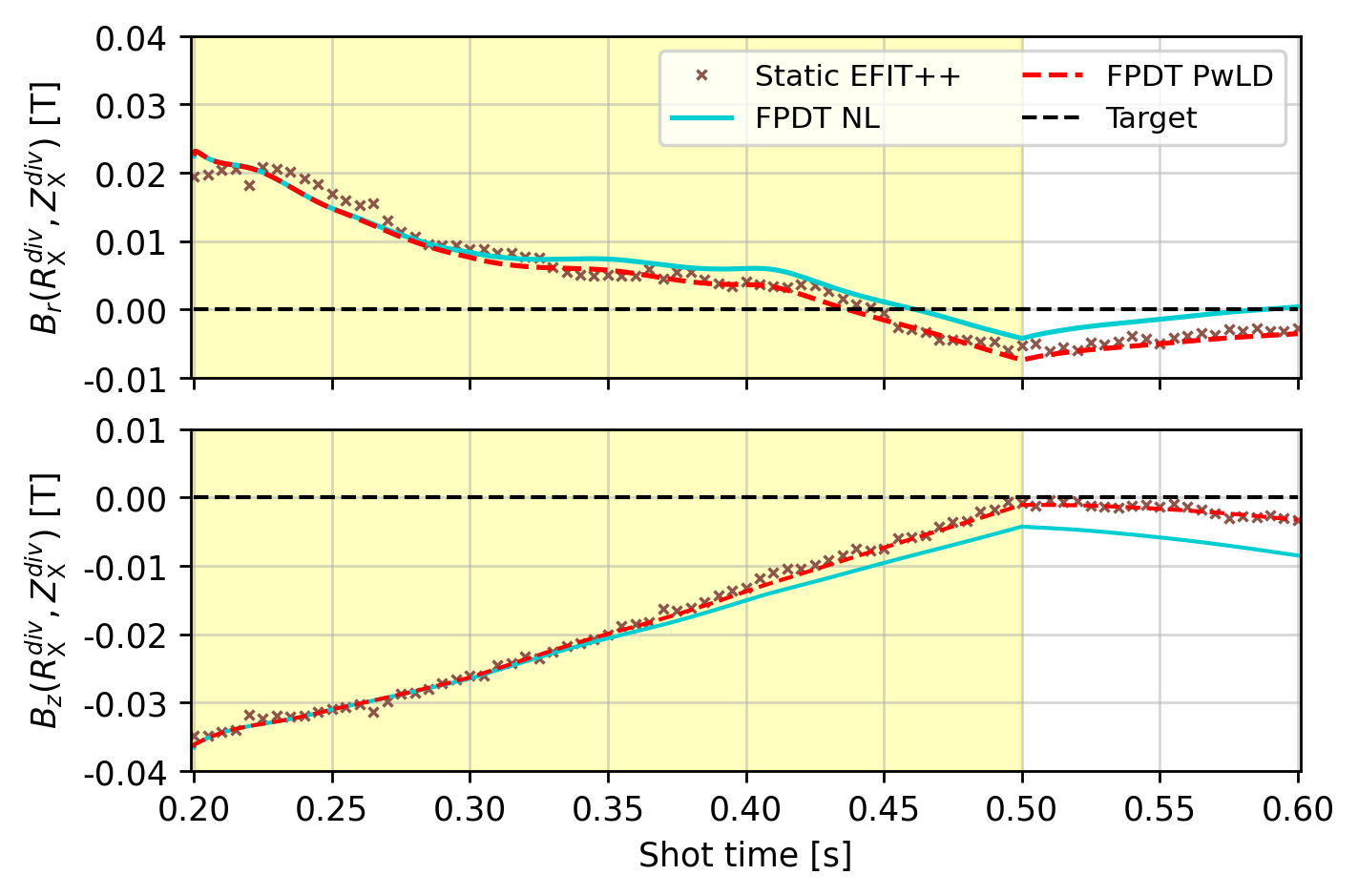}
    \end{subfigure}
    \caption{
    FPDT-simulated evolution of the radial (top) and vertical (bottom) magnetic fields at the position of the desired secondary X-point (in the lower divertor chamber) in MAST-U shot $50366$ using the NL (solid light blue) and PwLD (dashed red) simulation modes.
    Also shown are the target values (dashed black) and the EFIT\texttt{++} reconstructed values (brown).
    See the main text for a description of the background shading convention.
    }
    \label{fig:50366_BrBz}
\end{figure}

\begin{figure}[t!]
    \begin{subfigure}{0.9\linewidth}
        \centering
        \includegraphics[width=0.95\textwidth]{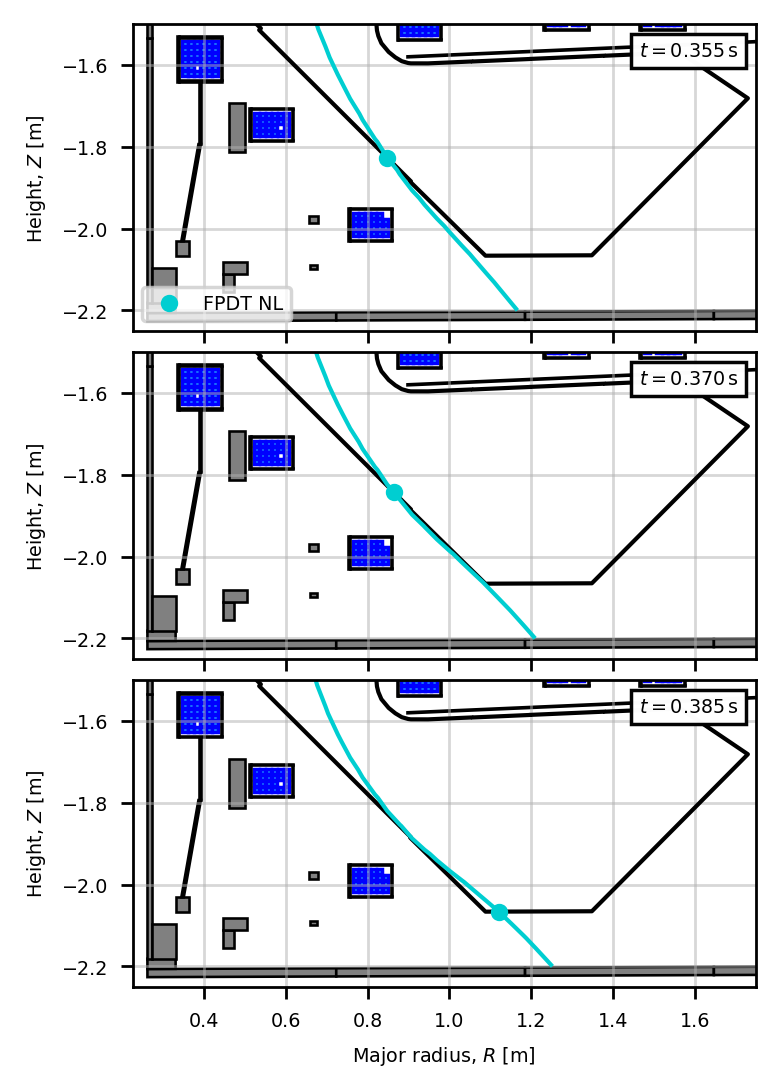}
    \end{subfigure}
    \caption{
    FPDT-simulated evolution of the divertor leg in MAST-U shot $50366$ using the NL (solid light blue) simulation mode.
    The dot indicates the location of the corresponding strikepoint $R_{\text{s}}$.
    }
    \label{fig:50366_strikepoint}
\end{figure}

As mentioned, this shot aimed to create a secondary X-point in the divertor chamber. 
The location could not, however, be directly controlled in FB, as it is not measured in real-time by LEMUR. 
To circumvent this limit, an additional FF drive is used, as shown in \cref{fig:50366_BrBz}. 
This drive is active between \SI{200}{\milli\second} and \SI{500}{\milli\second} and uses a specifically designed VC to form the X-point at $(R_{\text{X}^{\text{div}}}, Z_{\text{X}^{\text{div}}}) = (1.2, -1.8)$. 
This drive aims to form a magnetic null at this location, i.e. $B_r(R_{\text{X}^{\text{div}}}, Z_{\text{X}^{\text{div}}}) = B_z(R_{\text{X}^{\text{div}}}, Z_{\text{X}^{\text{div}}}) = 0$, while also fixing a point along the divertor leg in place, which explains the dynamics of the leg seen in \cref{fig:50366_strikepoint}).
In \cref{fig:50366_BrBz}, we show the evolution of magnetic null at this secondary X-point in both the NL and PwLD FPDT simulations, compared to the static EFIT\texttt{++}-based equilibria.
Both simulations approach the target values toward the end of the FF drive period.
We note that no PwL (dotted dark blue) simulation is shown, as this additional X-point was not included in the list of plasma descriptors.

\begin{figure*}[t!]
    \begin{subfigure}{0.99\linewidth}
        \centering
        \includegraphics[width=0.99\textwidth]{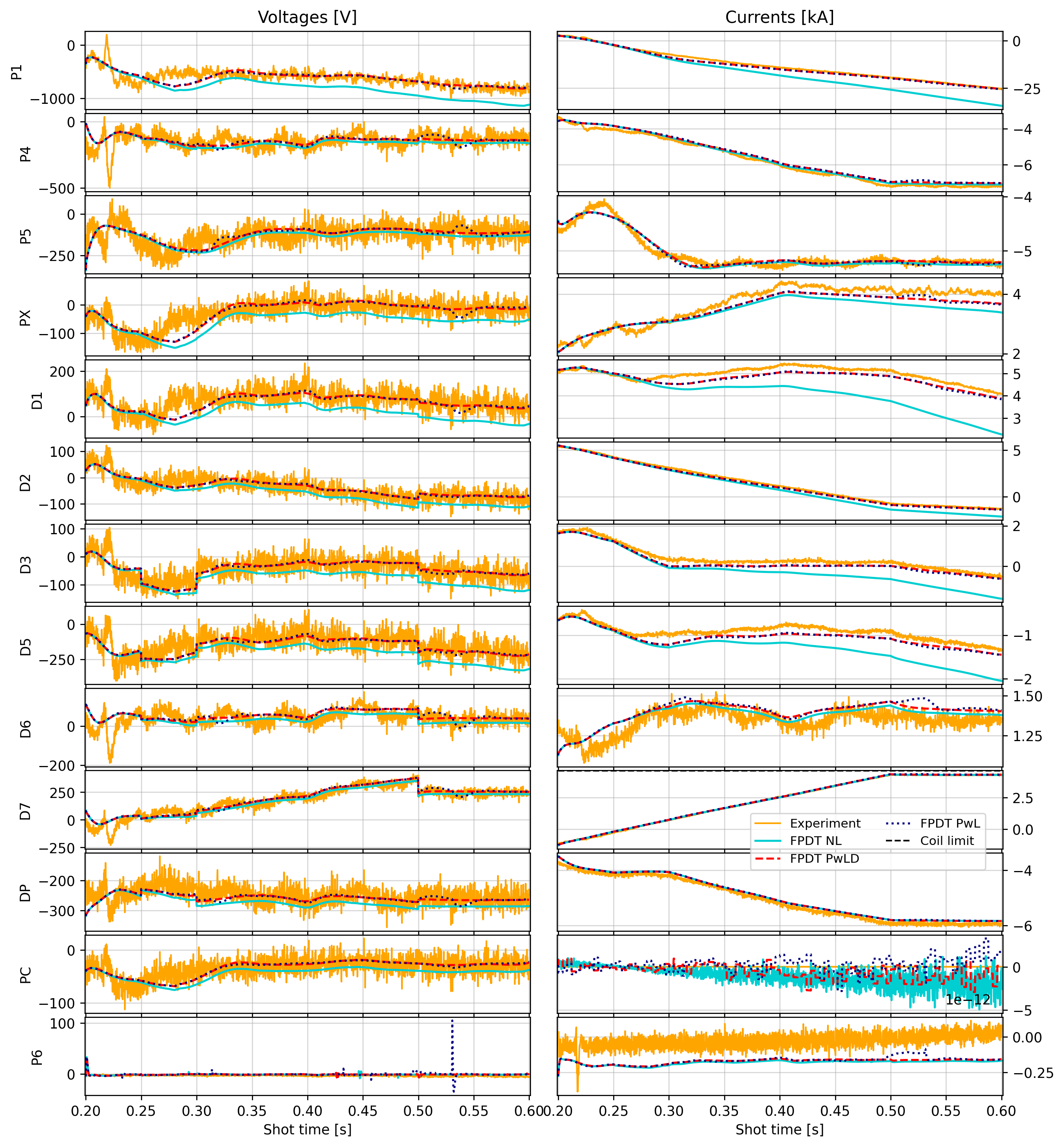}
    \end{subfigure}
    \caption{
    FPDT-simulated evolution of voltages (left panels) and currents (right panels) in all thirteen active PF coils (top to bottom) for MAST-U shot $50366$ using the NL (solid light blue), PwLD (dashed red), and PwL (dotted dark blue) simulation modes.
    Also shown in some panels are the coil voltage and current limits (dashed black) alongside the measured experimental values from the discharge (orange).
    }
    \label{fig:50366_currents}
\end{figure*}

The evolution of the coil voltages and currents is shown in \cref{fig:50366_currents}, with overall good agreement with experiment.
Deviations of some coil currents from the measurements again indicate that further improvements to the resistance and inductance matrices are possible.

The simulation execution times are similar to those of the previous experiment but we report them here for completeness. 
The NL simulation took $74\,$min \SI{59}{\second}; the PwLD took $6\,$min \SI{4}{\second} (with $8$ relinearisations); and the PwL took $2\,$min \SI{37}{\second} (with $16$ relinearisations).
Note that some of the PwL linearisation thresholds were adjusted to account for the larger shape changes: $Z_{\text{X}}$ was increased to \SI{2.5}{\centi\meter}, $R_{\text{nose}}$ reduced to \SI{1}{\centi\meter}, and $R_{\text{s}}$ reduced to \SI{5}{\centi\meter}.


\section{Summary and outlook} \label{sec:discussion}

This paper presents the FPDT, a simulation framework for axisymmetric tokamak plasma equilibria that couples FreeGSNKE's evolutive equilibrium solver with a new customisable virtual PCS class. 
While the virtual PCS presented here reflects the MAST-U control architecture, the modular design of the FPDT means that alternative magnetic control strategies such as isoflux or gap control can be implemented by substituting the relevant controller components without modifying the underlying equilibrium solver or communication infrastructure. 

We validate the efficacy of the FPDT by re-simulating existing MAST-U shots with advanced divertor configurations, using the same feedback/feedforward reference waveforms and control settings as in the real discharge.
The simulations show excellent quantitative agreement with both reference waveforms and experimental measurements from the discharge itself. 
The agreement observed in the evolved coil currents is particularly noteworthy, as any discrepancy between measured data and the simulations would be expected to accumulate over time and manifest itself as a progressively increasing deviation.
Small divergences in the coil current trajectories are observed nonetheless, and are likely due to small inaccuracies in the machine calibration---something to investigate in future work. 


Three different simulation modes were compared: NL, PwLD, and PwL, with the latter two capable of automatically updating their respective linearisations of the dynamics (and GS for PwL) as a simulation progresses.
We show, however, that both the PwLD and PwL modes can produce trajectories with accuracy comparable to the NL mode, while reducing computational cost by approximately an order of magnitude.
Using either of these piecewise linear modes enables the rapid evolution of key plasma parameters, with turnaround times of only a few minutes, making them well suited to inter-shot plasma scenario development and control design.

The validation demonstrated here ensures the FPDT is ready for use in predictive shape control modelling of the flat-top phase, enabling pre-shot testing of challenging plasma configurations. 
These predictions do, however, remain dependent on the reconstructed plasma profiles from previous discharges---something that could be avoided with the addition of transport and current diffusion models in the future. 
Beyond classical control studies, the FPDT also provides a powerful framework for the development and testing of AI-supported control schemes such as those based on emulated VCs \citep{agnello2024, cavestany2025, pentland2026}.

Future work will focus on the use of the FPDT for pre-shot control validation in the fifth MAST-U campaign and extending the simulation window to include the ramp-up phase---where vessel currents play a larger role in the equilibrium evolution. 
This would enable full shot scenario and control design, linking the plasma and pre-magnetisation phases together. 
In addition, a dedicated power supply model that incorporates delays or nonlinearities in the dynamics of power supply switching would be beneficial.

\section*{Acknowledgements}

We would like to thank Geof Cunningham, Stuart Henderson, James Harrison, Jimmy Measures, and Oliver Bardsley (UKAEA) for some very insightful and helpful discussions around the MAST-U PCS.


This work was funded by the Fusion Computing Lab collaboration (between UKAEA and STFC Hartree) and part funded by the EPSRC Energy Programme (EP/W006839/1).

For the purpose of open access, the authors have applied a Creative Commons Attribution (CC BY) licence to any author accepted manuscript version arising from this submission.
To obtain further information, please contact publicationsmanager@ukaea.uk.


\section*{Data availability}

The FPDT can be found within the FreeGSNKE open-source suite at \url{https://github.com/FusionComputingLab/freegsnke}.




\section*{Declarations}
The authors have no conflicts of interest to declare.


\begin{appendices} \label{appendix}
\crefalias{section}{appendix}

\section{Piecewise linear evolution for the dynamics and GS} \label{app:PwLD}

In the PwL simulation mode, the GS equation \eqref{eq:GS_PDE} is not solved at each time step.
Instead, the plasma state is provided by a reduced-order representation comprising a set of plasma descriptors most relevant to control.
In this paper, the user-defined plasma descriptors, denoted $\bm{D}(t)$, consist of variables that describe the plasma's local or global shape (e.g. midplane radii, elongation, etc.) or position (e.g. vertical position).

Following the solution of \eqref{eq:circuit_eqs}-\eqref{eq:plasma_eq} at a given time $t$ in PwL mode, the plasma descriptors are evolved according to 
\begin{align*}
    \bm{D}(t) = \bm{D}(t_0) + \delta_{\bm{I}_e}(t) \left. \frac{\partial \bm{D}}{\partial \bm{I}_e} \right|_{t=t_0} 
    + \delta_{\bm{\theta}}(t) \left. \frac{\partial \bm{D}}{\partial \bm{\theta}} \right|_{t=t_0},
\end{align*}
where $\delta_x(t) = x(t) - x(t_0)$ and $\bm{I}_e = (\bm{I}_m, I_p)$.
The Jacobians of the descriptors with respect to $\bm{I}_e$ and $\bm{\theta}$ are calculated at the same time $t_0$ as the Jacobians used to solve the dynamics \eqref{eq:circuit_eqs}-\eqref{eq:plasma_eq}.
This is done using the equilibrium at time $t_0$, which triggered the latest relinearisation process.
This enables faster simulations that bypass the large number of GS solves of the PwLD mode, while retaining a physically meaningful description of plasma evolution, thereby facilitating controller development and testing.


\end{appendices}



\begingroup
\small                        
\bibliographystyle{abbrvnat}  
\bibliography{references}  
\endgroup


    
\end{document}